\documentclass[11pt,preprintnumbers,aps,amssymb,nofootinbib,amsmath]
{revtex4}
\usepackage{epsfig,epsf}
\usepackage{xfrac} % makes nice fractions with \sfrac{ }{ } command (in or out of Math mode)
\usepackage{bm} % puts greek and math symbols in boldface using \bm
\newcommand{\beq}{\begin{equation}}
\newcommand{\beql}[1]{\begin{equation}\label{#1}}
\newcommand{\eeq}{\end{equation}}
\newcommand{\eq}[1]{(\ref{#1})}
\newcommand{\fig}[1]{Fig.~\ref{#1}}
\renewcommand{\sec}[1]{Sec.~\ref{#1}}
\newcounter{topiccounter}
\renewcommand{\b}[1]{{\bm #1}} % puts roman letter into streight bold 
\newcommand{\unit}[1]{\hat {{\bf #1}}} % unit vector

\newcommand{\as}{\alpha_s}
\newcommand{\bas}{\bar\alpha_s}

\newcommand{\Pom}{\text{I\!P}}
\newcommand{\Q}{\mathcal{Q}}
\begin{document}

\title{Properties of inclusive hadron production in  Deep Inelastic Scattering on heavy nuclei at low-$x$}

\author{Kirill Tuchin and Dajing Wu\\}

\affiliation{Department of Physics and Astronomy, Iowa State University, Ames, IA 50011}

\date{\today}

\pacs{}

\begin{abstract}

In this paper we present a comprehensive study of  inclusive hadron production in DIS at low $x$. Properties of the hadron spectrum are different in different kinematic regions formed by three relevant momentum scales: photon virtuality $Q^2$, hadron transverse momentum $k_T$ and  the saturation momentum $Q_s(x)$. We investigate each kinematic region and derive the corresponding asymptotic formulas for the cross section at the leading logarithmic order. We also analyze the next-leading-order (NLO) corrections to the BFKL kernel that are responsible for the momentum conservation.
In particular, we establish the asymptotic behavior of the forward elastic dipole--nucleus scattering amplitude at high energies deeply in the saturation regime and a modification of the pomeron intercept. We study the nuclear effect on the inclusive cross section using the nuclear modification factor and its logarithmic derivative. We argue that the later is proportional  to the difference between the anomalous dimension of the gluon distribution in nucleus and in proton and thus is a direct measure of the coherence effects.  To augment our arguments and present quantitative results we performed numerical calculations in the kinematic region that may be accessible by the future DIS experiments. 

\end{abstract}

\maketitle

%%%%%%%%%%%%%%%%%%%%%%%%%%%%%%%%%%%%%%%%
\section{Introduction}\label{sec:intr}

In the last decade we have learned a great deal about  gluon saturation/color glass condensate \cite{Gribov:1984tu,LR,Mueller:1985wy,Mueller:1993rr,Mueller:1994jq,Mueller:1994gb,McLerran:1993ka,McLerran:1993ni,McLerran:1994vd,Kovchegov:1996ty,Kovchegov:1997pc,JalilianMarian:1996xn,JalilianMarian:1997jx,JalilianMarian:1997gr,JalilianMarian:1997dw,JalilianMarian:1998cb,Kovner:2000pt,Weigert:2000gi,Iancu:2000hn,Ferreiro:2001qy,Kovchegov:1999yj,Kovchegov:1999ua,Balitsky:1995ub,Balitsky:1997mk,Balitsky:1998ya,Iancu:2003xm,JalilianMarian:2005jf}  thanks to the relativistic $dAu$ and $AuAu$ program at RHIC. The future DIS programs at EIC and LHeC promise to provide even more detailed information about structure of the nuclear matter at low $x$. How successful that program will be depends a lot on our ability to pinpoint the processes that are most sensitive to the low-$x$ regime. In this paper we study one such process -- inclusive hadron production in $eA$ scattering. It has been a subject of intense theoretical investigation over the past decade  \cite{Kovchegov:1998bi,Kovchegov:2001sc,Braun:2000bh,Dumitru:2001ux,Blaizot:2004wu,Kharzeev:2002pc,Kharzeev:2003wz,Kharzeev:2004yx,Baier:2003hr,Iancu:2004bx}  and has proved to be a  powerful tool in $dA$ collisions at RHIC. On the one hand, we expect that $p(d)A$ and $eA$ processes have very much in common due to the Pomerantchuk theorem, that states that all high energy scattering processes are mediated by exchange of a collective gluon state -- known as pomeron -- that has vacuum quantum numbers. On the other hand, proton wave function is characterized by a soft, non-perturbative scale, whereas the virtual photon wave function can be calculated using the perturbation theory and is characterized by virtuality $Q^2$. A possibility to dial  $Q^2$ is a  great advantage of DIS. Our main goal in this paper is to provide a thorough 
analysis of the inclusive hadron production in various kinematic regions characterized by three dimensional scales: photon virtuality $Q^2$, hadron momentum $k_T$ and the saturation momentum $Q_s$ and to produce numerical predictions for both novel and well-known quantities that can be tested at EIC and/or LHeC.

Our paper is organized as follows. In  \sec{sec:dipoletoDIS} we use the dipole model \cite{Mueller:1989st} to relate the DIS $\gamma^*A$ cross section to that of the  color dipole $q\bar q+A$.  The $\gamma^*A$ differential cross section can be  expressed in a factorized form as a product of the  light-cone wave function of the virtual photon $\gamma^*$  and $q \bar q + A$ differential cross section. In \sec{sec:logs} we review the properties of the BFKL pomeron \cite{Kuraev:1977fs,Balitsky:1978ic} and the unintegrated gluon distribution function at LO, particularly we emphasize the leading logarithmic asymptotics. These are used in \sec{sec:dipole} to derive the asymptotic properties of gluon production in dipole--nucleus scattering in various kinematic regions. In \sec{sec:assym} the result is further generalized to the case of LO gluon production in DIS.

The NLO corrections to the inclusive hadron production are rather complex.  These include NLO correction to the BFKL kernel  \cite{Fadin:1998py,Ciafaloni:1998gs}, \cite{Ciafaloni:2003rd,Ciafaloni:2002xk,Ciafaloni:2002xf,Forshaw:2000hv,Ciafaloni:2001db,Brodsky:1998kn,Ross:1998xw,Levin:1998pka,Armesto:1998gt,Kovchegov:1998ae}, running coupling corrections \cite{Levin:1994di,Mueller:2002zm,Triantafyllopoulos:2002nz,Braun:1994mw,Kovchegov:2006vj,Balitsky:2006wa,Kovchegov:2006wf,Kovchegov:2007vf}  and  energy conservation \cite{Kuokkanen:2011,Weigert:2007hk,Chachamis:2004ab} corrections to BK \cite{Kovchegov:1999yj,Kovchegov:1999ua,Balitsky:1995ub,Balitsky:1997mk}. It has been argued in \cite{Kormilitzin:2010at} that  energy conservation is the most important phenomenological effect beyond the LO. Therefore, in \sec{sec:nlo} we investigate the role of this effect on inclusive hadron production.  In our calculations we rely on a phenomenological approach suggested in \cite{Kormilitzin:2010at,Gotsman:2004xb} where a modified BK (mBK) equation that satisfies energy  conservation was derived. It was utilized in \cite{Gotsman:2002yy,Kuokkanen:2011} to calculate the NLO corrections to the  total DIS cross section. mBK equation serves as the basis for our NLO calculations.  First, we derive  the dipole scattering amplitude in dilute and saturation regimes;  the corresponding expressions are given by \eq{N-nlo-dil} and \eq{s-sat-1} respectively.  We argue that the energy  conservation effects decrease the energy dependence of the saturation momentum. These results are used 
for  computation of dipole density in various asymptotic regimes. Similarly to our analysis of LO case, we explore the NLO gluon production first for dipole---nucleus process and then for DIS scattering. 

It is very instructive to know how the DIS on a heavy nucleus is different from DIS on a proton at low $x$. Had the coherence length been short, of the order of the proton radius, the hadron production  in $\gamma^*A$ would have been equal the incoherent sum of $A$ $\gamma^*N$ processes. However, since the coherence length is larger than the nuclear radius, the entire process  is coherent. Because it is interesting to compare the coherent and incoherent regimes, one introduces the nuclear modification factor (NMF) $R$ that calibrates the cross section in $\gamma^*A$ with that of $\gamma^*N$  rescaled by atomic weight $A$. \sec{sec:nmf} is devoted to the study of the  properties of this quantity as a function of the hadron transverse momentum, photon virtuality and atomic weight.  

We expect that at EIC/LHeC kinematic region the low-$x$ evolution effects start to play an important role rendering the anomalous dimensions dependent on atomic weight. This manifests itself in inclusive hadron production in $dA$ collisions at RHIC as the transition from the Cronin enhancement at mid-rapidity to suppression of the NMF at forward rapidities even at $k_T>Q_s$. In order to evaluate how steep is the dependence  of the NMF on rapidity, we introduce a new observable $J$, defined as the logarithmic derivative of $R$, viz.\ $d\ln R/dy$. We demonstrate in \sec{sec:nmf} that at $k_T\gg Q_s$, $J$ is proportional to the difference of the anomalous dimensions of the gluon distribution in nucleus and in proton. Without the low-$x$ evolution one expect $J$ to vanish. However, due to the low-$x$ evolution $J$ acquires a finite negative value. Therefore, $J$ can serve as a direct probe of the effect of the slow-$x$ evolution on the nuclear gluon distribution function.

The numerical computations are presented in \sec{sec:numer}. We use the bCGC model \cite{Kowalski:2006hc} for the dipole-nucleus forward scattering amplitude, albeit with the simplified $b$-dependence. In \fig{fig:F2} we plot   $d^2F_2/d\ln k_T^2\,dy$ as a function of photon virtuality $Q^2$ and  hadron transverse momentum $k_T$ and rapidity $y=\ln(1/x_\Pom)$.\footnote{We use the $x_\Pom$ notation borrowed from the diffractive DIS where it denotes the momentum fraction carried by the pomeron. It does not have this simple interpretation in our case because the interaction is inelastic.} In order to emphasize the role played by the 
 NLO effects we exhibit both LO and NLO results in each plot for the structure function. 
 In \fig{fig:F2} we see that the NLO calculation yields much smaller cross section for inclusive hadron production than the LO one. Additionally, its functional dependence on $k_T$, $Q^2$ and $y$ is substantially weaker in NLO than in LO. This  is in accordance with our observation in \sec{sec:nlo} that NLO correction reduces the anomalous dimension of the gluon distribution. Interestingly, most of the NLO effect cancels in the NMF which appears to be a robust quantity in this respect. This indicates that the energy conservation effect factors out to a large extent from the inclusive cross section. 
 
The NMF shown in  \fig{fig:NMF} displays a number of interesting features. First, the NMF is strongly suppressed at small $k_T$'s but exhibits 
 an enhancement toward higher $k_T$'s where the Cronin effect ($R>1$) is observed. 
 This seems to be in contrast with  $pA$ collisions \cite{Kharzeev:2003wz} where the Cronin effect gives way to suppression of NMF at all $k_T$'s as the hadron rapidity  increases. This is the result of the linear evolution in the rapidity interval between the virtual photon and the hadron. This evolution produces dipoles of different sizes that scatter in the nucleus with different amplitudes. At small $k_T$ large dipoles, on which the gluon saturation effects are stronger, dominate the cross section, whereas at higher $k_T$ smaller dipoles contribute to the NMF enhancement. Second, we observe a relatively weak $A$-dependence. This is also a result of the averaging over different dipoles. Third, we note a peculiar $Q^2$ dependence that is explained in \sec{sec:numer}. 
 
To investigate the rapidity dependence in more detail we plot the logarithmic slope of the nuclear modification factor  $J$ on \fig{fig:Jqqbar} (for dipole-nucleus scattering). We see that it is negative for the entire kinematic region indicating the graduate suppression of the NMF towards large rapidities. This is in agreement with our arguments in \sec{sec:nmf}. We argue that  $J$ is directly proportional to the difference between the anomalous dimensions of the gluon distribution function in the nucleus and in proton.  Hence we believe that measuring $J$ is a great tool for exploring the low-$x$ regime of QCD.

We summarize our results in \sec{sec:summary}.

%%%%%%%%%%%%%%%%%%%%%%%%%%%%%%%%%%%%%%%%
\section{From $\gamma^*A$ to $q\bar q +A$ scattering}\label{sec:dipoletoDIS}

The dominant contribution to the inclusive hadron production in DIS at low-$x$, at rapidities away from the virtual photon and nucleus fragmentation regions,  comes from the fragmentation of fast $s$-channel gluons \cite{LR}. The cross section for inclusive production of a gluon of transverse momentum $\b k$ at rapidity $y$ in deep inelastic scattering can be represented as an integral in the configuration space  \cite{Nikolaev:1990ja}:\footnote{We use the notation $\b k^2= k^2= k_T^2$, where $\b k$ is a vector transverse to the collision axis.}
\beql{xsect1}
\frac{d\sigma^{\gamma^*A}(\b k,y;Q)}{d^2kdy}= \frac{1}{2\pi^2}\int d^2 r \int_0^1 dz\, \Phi(\b r,z,Q)\,
\frac{d\sigma^{q\bar q+A}(\b k, y; \b r)}{d^2kdy}\,,
\eeq
where the virtual photon wave function $\Phi$ describes splitting of a photon of virtuality $Q^2$ into $q\bar q$ color dipole. It is given by  
\begin{subequations}\label{w-f}
\begin{align}
\Phi(\b r,z)(Q, r,z)&= |\Psi^{\gamma^*}_T(Q, r,z)|^2+|\Psi^{\gamma^*}_L(Q, r,z)|^2\,,\\
\Phi_T(\b r,z)(Q, r,z)&=2N_c\sum_f\frac{\alpha_\text{em}^f}{\pi}\{a^2 K_1^2( r a)[z^2+(1-z)^2]+m_f^2 K_0^2( r a)\}\,,\\
\Phi_L(\b r,z)(Q, r,z)&=2N_c\sum_f\frac{\alpha_\text{em}^f}{\pi} 4Q^2z^2(1-z)^2K_0^2( r a)\,.
\end{align}
\end{subequations}
Here $a^2= Q^2z(1-z)+m_f^2$, $\alpha_\text{em}^2= e^2 z_f^2/(4\pi)$, with $z_f$ being electric charge of quark $f$ in the units of electron charge $e$.  The cross section for inclusive gluon production in dipole--nucleus scattering reads \cite{Kovchegov:2001sc}
\beql{xdip1}
\frac{d\sigma^{q\bar q+A}(\b k, y; \b r)}{d^2kdy}=\frac{2\as C_F}{\pi^2}\frac{1}{k^2}\int d^2b \int d^2r' e^{-i\b k\cdot \b r'}[\nabla^2_{r'}N_G(\b r',\b b',y)]\,[\nabla^{-2}_{r'}n(\b r, \b r', Y-y)]\,,
\eeq
Here the dipole density $n( r, r', Y-y)d^2r'$ is the number of daughter dipoles of size $r'$ in the interval $d^2r'$ produced by a parent dipole of size $r$ at the relative rapidity $Y-y$ \cite{Mueller:1993rr,Mueller:1994jq,Mueller:1994gb}.  It satisfies the BFKL equation \cite{Kuraev:1977fs,Balitsky:1978ic} with the initial condition
\beql{bfkl-ic}
n(\b r,\b r',0)=\delta(\b r-\b r')\,.
\eeq
At the leading logarithmic order, the corresponding solution is \cite{Kuraev:1977fs,Balitsky:1978ic}
\beql{n-sol}
n(\b r, \b r', y)= \frac{1}{2\pi^2 r'^2}\int_{-\infty}^\infty d\nu\, e^{2\bas \chi(\nu)y}\left(\frac{r}{r'}\right)^{1+2i\nu}\,
\eeq
with the eigevalue function $\chi$ given by
\beql{chi}
\chi(\nu)=\psi(1)-\frac{1}{2}\psi\left(\frac{1}{2}-i\nu\right)-\frac{1}{2}\psi\left(\frac{1}{2}+i\nu\right)\,,
\eeq
where $\psi(\nu)=\Gamma'(\nu)/\Gamma(\nu)$.

Let $f(\b r, \b r', y)$ be the particular solution of the two-dimensional Poisson equation
\beql{eq1}
\nabla^2_{r'} f(\b r, \b r',y)= n(\b r, \b r', y)\,.
\eeq 
Employing \eq{n-sol} we derive the Melin representation of $f$
\beql{f-sol}
f(\b r, \b r', y)=\nabla^{-2}_{r'}n(\b r, \b r', y)= \frac{1}{2\pi^2}\int_{-\infty}^\infty d\nu\, \frac{1}{(2i\nu+1)^2}\, e^{2\bas \chi(\nu)y}\left(\frac{r}{r'}\right)^{1+2i\nu}\,.
\eeq
It is convenient to write \eq{xdip1} as a convolution in the momentum space. To this end we 
introduce the Fourier-image of $f$ with respect to the second argument:
\beql{tilde-f}
\tilde f(\b r, \b q, y)= \int d^2r'\, e^{-i\b q\cdot \b r'}\,f(\b r, \b r', y)=
\frac{r}{\pi q}\int_{-\infty}^\infty d\nu\, e^{2\bas \chi(\nu)y}\left(\frac{rq}{2}\right)^{2i \nu}\frac{\Gamma\left(\frac{1}{2}-i\nu\right)}{\Gamma\left(\frac{1}{2}+i\nu\right)(2i\nu+1)^2}
\eeq
and the unintegrated gluon distribution function of the nucleus \cite{LR,Kovchegov:2001sc}
\beql{dist-def}
\varphi_A(\b k, y)=\frac{C_F}{\as(2\pi)^3}\int d^2b\int d^2r \, e^{-i\b k\cdot\b r}\,\nabla_r^2N_G(\b r,  \b b,y)\,.
\eeq 
$N_G(\b r,  \b b,y)$ is the forward scattering amplitude of a color gluon (or adjoint) dipole $\b r$ on the nucleus at impact parameter $\b b$ at the relative rapidity $y$. It obeys the BK equation \cite{Balitsky:1995ub,Kovchegov:1999yj} and its  properties are discussed in the next section.
Using \eq{tilde-f} and \eq{dist-def}  in \eq{xdip1} we get
\beql{xdip3}
\frac{d\sigma^{q\bar q+A}(\b k, y; \b r)}{d^2kdy}=\frac{4\as^2}{\pi k^2} \int d^2p\, \varphi_A(\b p, y)\,
\tilde f(\b r, \b k-\b p, Y-y)\,.
\eeq

\section{Logarithmic approximations}\label{sec:logs}

%%%%
\subsection{Asymptotic expressions  for $\tilde f$}\label{sec:ass-f}

It is worthwhile to list here the asymptotic formulas for $\tilde f$ in various kinematic regions (we follow notations of  \cite{Li:2008bm,Li:2008jz,Li:2008se} were more details can be found).
\noindent
\begin{enumerate}

\item $\as y\gg \ln^2\frac{rq}{2}$. In this case the eigenfunction \eq{chi} can be expanded near its minimum $\chi\approx 2\ln 2-7\zeta(3)\nu^2$.  Expression under the $\nu$-integral in \eq{f-sol} has a saddle point   at 
\beq\label{sp2}
i\nu_\text{sp} = \frac{\ln(2/rq)}{14\zeta(3)\bas y} \,.
\eeq
In this approximation integration over $\nu$ in \eq{f-sol} produces 
\beql{f-dif}
\tilde f(\b r, \b q, y)=\frac{r}{q}\frac{1}{\sqrt{14\pi\zeta(3)\bas\, y}}\, e^{(\alpha_P^{(0)}-1)y}\, e^{-\frac{\ln^2\frac{rq }{2}}{14\zeta(3)\bas\, y}}\,,
 \eeq
with $\alpha_P^{(0)}-1=4\bas \ln 2$.

\item $rq<2$ and $\ln\frac{2}{rq}\gg \as y$. In this region, the leading contribution to the $\nu$-integral stems from  the  pole at $i\nu = 1/2$. Approximating  the eigenfunction as $\chi\approx 1/(1-2i\nu)$ and employing the  saddle point method in \eq{f-sol} again  yields
\beql{f-dlog-1}
\tilde f(\b r, \b q, y)=\frac{r^2}{8\sqrt{\pi}}\frac{1}{\left(2\bas y \ln\frac{2}{rq} \right)^{1/4}} e^{2\sqrt{2\bas y \ln \frac{2}{rq}}}\,.
\eeq
The saddle point is 
\beq\label{spp}
2i\nu_\text{sp}= 1-\sqrt{\frac{2\bas y}{\ln\frac{2}{rq}}}\,.
\eeq

\item $rq>2$ and $\ln\frac{rq}{2}\gg \as y$. Now,  another pole in $\chi$ dominates, $\chi\approx 1/(1+2i\nu)$ with the result for $\tilde f$
\beql{f-dlog-2}
\tilde f(\b r, \b q, y)=\frac{1}{2q^2\sqrt{\pi}}\frac{1}{\left(2\bas y \ln\frac{rq}{2} \right)^{1/4}} e^{2\sqrt{2\bas y \ln \frac{rq}{2}}}\,
\eeq
 and for the saddle point
\beq\label{spp2}
 2i\nu_\text{sp}= -1+\sqrt{\frac{2\bas y}{\ln\frac{rq}{2}}}\,.
\eeq
 
\end{enumerate}

%%% 
\subsection{Properties of $\varphi_A$}

Unintegrated gluon distribution $\varphi_A$ is defined by \eq{dist-def}. $N_G(\b r, y,\b b)$ stands for the forward elastic gluon dipole scattering amplitude. At large $N_c$,  the gluon dipole is equivalent to two $q\bar q$ dipoles each of which scatters with  amplitude $N(\b r, y, \b b)$. Therefore,
\beql{Ng-N} 
N_G(\b r, \b b,y)=2N(\b r, \b b,y)-N^2(\b r, \b b,y)
\eeq
The $q\bar q$ scattering amplitude satisfies the BK equation \cite{Balitsky:1995ub,Kovchegov:1999yj} and its properties are well-known.
Initial condition for the BK equation is the Glauber-Mueller formula \cite{Mueller:1989st} for the forward scattering amplitude $N$ of a $q\bar q$ color dipole on the nucleus:
\beql{init-bk}
N(\b r, \b b,0)=1-e^{-\frac{1}{8}\b r^2 Q_{s0}^2}\,.
\eeq
The \emph{gluon} saturation momentum \cite{Gribov:1984tu} at initial rapidity $y=0$, which corresponds to the Bjorken variable $x_0$ such that $y=\ln\frac{x_0}{x}$, is related to gluon distribution function $xG$ at $x=x_0$ as 
\beq\label{Qsat}
Q_{s0}^2=\frac{4\pi^2\as N_c}{N_c^2-1}\,\rho\, T(\b b)\, x_0G(x_0,1/\b r^2)\,,
\eeq
where $\rho$ is the nuclear density, $T(\b b)$ is the nuclear thickness function as a function of the impact parameter $\b b$. The gluon distribution function at the leading order in $\as$, i.e.\ in the two-gluon exchange approximation, reads
\beq\label{xG}
xG(x,1/\b r^2)=\frac{\as C_F}{\pi}\ln \frac{1}{\b r^2 \Lambda^2}\,,
\eeq
with $\Lambda$ being some non-perturbative momentum scale characterizing the nucleon's wave function. 
Using \eq{init-bk} in \eq{Ng-N} we derive the initial condition for the gluon dipole scattering amplitude
\beql{init-bk-ng}
N_G(\b r, \b b,0)=1-e^{-\frac{1}{4}\b r^2 Q_{s0}^2}\,.
\eeq

Let us now list some properties of the amplitude $N_G$, see \cite{Li:2008bm,Kharzeev:2003wz} for details.
\noindent
\begin{enumerate}
\item  At $r\ll 1/Q_{s0}$ the BK equation reduces to the BFKL equation, which must be solved with the initial condition  $N(\b r, \b b, 0)\approx r^2Q_{s0}^2/4$. Small dipoles scatter independently, perforce $N_G\approx 2N$. Thus, in this region 
\beql{N-lin}
N_G(\b r, \b b, y)= \int_{-\infty}^\infty d\nu\, e^{2\bas\chi(\nu)y}\, (rQ_{s0})^{1+2i\nu}\,\frac{1}{8\pi}\frac{1+(1-2i\nu)\ln\frac{Q_{s0}}{\Lambda}}{(1-2i\nu)^2}\,.
\eeq

\item In particular, if $r\ll 1/Q_{s0}$ and $\ln \frac{1}{rQ_{s0}}\gg \as y$ the solution is
\beql{N-lin-dif}
N_G(\b r, \b b, y)=\frac{\sqrt{\pi}}{8\pi}\frac{(\ln\frac{1}{rQ_{s0}})^{1/4}}{(2\bas y)^{3/4}}r^2Q_{s0}^2\left(1+\sqrt{\frac{2\bas y}{\ln\frac{1}{rQ_{s0}}}}\ln \frac{Q_{s0}}{\Lambda}\right)e^{2\sqrt{2\bas y \ln \frac{1}{rQ_{s0}}}}\,.
\eeq

\item For $r\ll 1/Q_{s0}$ and $\as y\gg \ln^2 \frac{1}{rQ_{s0}}$ we have
\beql{N-lin-dlo}
N_G(\b r, \b b,y)=\frac{rQ_{s0}}{4}\frac{\ln\frac{Q_{s0}}{\Lambda}}{\sqrt{14\zeta(3)\pi \bas y}}e^{(\alpha_P-1)y}\,e^{-\frac{\ln^2 (rQ_{s0})}{14\zeta(3)\bas y}}
\eeq

\item\label{is4} The saturation region is characterized by the saturation momentum $Q_s(y)$. With the double logarithmic accuracy it reads \cite{Levin:1999mw,Levin:2000mv,Levin:2001cv}
\beql{ss}
Q_s(y)=Q_{s0}e^{2\bas y}
\eeq
In the saturation region $r>1/Q_s$, solution to the BK equation is \cite{Levin:1999mw,Levin:2000mv,Levin:2001cv}
\beql{sol-sat}
N(\b r, \b b, y)= 1-S_0e^{-\frac{1}{8}\ln^2(r^2Q_s^2)}\,,
\eeq
where $S_0$ is a constant that can be determined by matching $N$ from \eq{sol-sat} with that of \eq{N-lin} at $r=2/Q_s(y)$. Consequently, 
\beql{sol-sat-ng}
N_G(\b r, \b b, y)= 1-S_0^2e^{-\ln^2(rQ_s)}\,,
\eeq
where we utilized \eq{Ng-N}.

\end{enumerate}

Eqs.~\eq{N-lin}-\eq{sol-sat-ng} are derived with the logarithmic accuracy. We can calculate $\varphi_A$ given by \eq{dist-def} in the same approximation as  
\beql{dist-app}
\varphi_A(\b k,y)\approx \frac{C_F}{\as(2\pi)^2}\int d^2b \int_0^{1/k} dr\, \frac{\partial}{\partial r}\left( r\frac{\partial}{\partial r}N_G(\b r, \b b, y)\right)=\frac{C_F}{\as(2\pi)^2k}\int d^2b\, \frac{\partial }{\partial r}N_G(\unit r/k,\b b, y)\,.
\eeq
We stress that this formula holds only in the asymptotic regions specified in 1-4 above; still this is a very useful approximation as it  captures the most essential features of the unintegrated gluon distribution. 

It is evident from \eq{dist-app}, that  in place of function $N_G(\b r, \b b, y)$ it is convenient to use function $\tilde N_G(k, b, y)= N_G(\unit r/k, \b b, y)$, where $\unit r= \b r/r$.  In particular, 
$\partial N_G(\unit r/k,  b, y)/\partial r = -k^2 \partial \tilde N_G(k, \b b, y)/\partial k$.\footnote{We assumed in  \eq{Qsat} that the $b$-dependence factors out in the initial condition; perforce it factors out in the solution for heavy nuclei. Therefore, scattering amplitudes depend only on the absolute value of vector $\b b$.} Plugging \eq{dist-app} into  \eq{xdip3} we obtain
\beql{xdip5}
\frac{d\sigma^{q\bar q+A}(\b k, y; \b r)}{d^2kdy}=\frac{\as C_F}{\pi^3 k^2} \int d^2b\int d^2p\, \frac{\partial \tilde N_G( p,  b, y)}{\partial\ln (1/p)}\,
\tilde f(\b r, \b p-\b k, Y-y)\,.
\eeq 

%%%%%
\section{Properties of the dipole--nucleus cross section}\label{sec:dipole}

 To calculate  the cross section for gluon production in dipole--nucleus scattering we need to evaluate the integral over the transverse momentum $\b p$ in the right-hand-side of \eq{xdip5}. It convenient to consider the inclusive cross section at a fixed  impact parameter $b$:
 \beql{def-g}
g(\b k,y,\b b;\b r)\equiv \frac{d\sigma^{q\bar q+A}(\b k, y; \b r)}{d^2kdy\,d^2b}\left( \frac{\as C_F}{\pi^3 k^2}\right)^{-1} =\int d^2p\, \frac{\partial \tilde N_G( p,  b, y)}{\partial\ln (1/p)}\,
\tilde f(\b r, \b p-\b k, Y-y)\,.
 \eeq
When taking the $p$-integral with the logarithmic accuracy in various kinematic regions it is useful to keep in mind that \eq{sol-sat-ng},\eq{N-lin}   imply that $\partial \tilde N_G/\partial\ln (1/p)\sim \ln(Q_s/p)\exp\{-\ln^2(Q_s/p)\}$ if $p\ll Q_s$ and $\partial \tilde N_G/\partial\ln (1/p)\sim Q_s^2/p^2$ if $p\gg Q_s$, while \eq{f-dlog-1},\eq{f-dlog-2} indicate that $\tilde f\sim 1/k^2$ if $k\gg 1/r$ and $\tilde f\sim r^2$, if $k\ll 1/r$.
\noindent
\begin{enumerate}
\item $k\gg Q_s\gg 2/r$. Due to the strong ordering of the relevant scales we have 
\beql{g-reg1}
g\approx 2\pi \int_{Q_s}^k dp p \frac{\partial \tilde N_G( p,  b, y)}{\partial\ln (1/p)}\,
\tilde f(\b r, \b k, Y-y)\,.
\eeq
Using \eq{N-lin} we derive
\begin{align}
\int_{Q_s}^k dp p \frac{\partial \tilde N_G( p,  b, y)}{\partial\ln (1/p)}\,
&= k^2\int_{-\infty}^\infty d\nu\, e^{2\bas\chi(\nu)y}\, \left(\frac{Q_{s0}}{k}\right)^{1+2i\nu}\,\frac{1}{8\pi}\frac{1+(1-2i\nu)\ln\frac{Q_{s0}}{\Lambda}}{(1-2i\nu)^2}\frac{1+2i\nu}{1-2i\nu}\nonumber \\
&\approx    \frac{\sqrt{\pi}}{8\pi}\frac{(\ln\frac{k}{Q_{s0}})^{3/4}}{(2\bas y)^{5/4}}Q_{s0}^2\left(1+\sqrt{\frac{2\bas y}{\ln\frac{k}{Q_{s0}}}}\ln \frac{Q_{s0}}{\Lambda}\right)e^{2\sqrt{2\bas y \ln \frac{k}{Q_{s0}}}}\,.
\label{g-reg1-A}
\end{align}
Thus, it follows upon substitution of \eq{f-dlog-2} and \eq{g-reg1-A} into \eq{g-reg1} and then into \eq{xdip5} that 
\begin{align}
\frac{d\sigma^{q\bar q+A}(\b k, y; \b r)}{d^2kdy}&=\frac{\as C_F}{8\pi^3 k^4} \int d^2b\, Q_{s0}^2\,\frac{(\ln\frac{k}{Q_{s0}})^{3/4}}{(2\bas y)^{5/4}(2\bas(Y-y)\ln\frac{kr}{2})^{1/4}}\left(1+\sqrt{\frac{2\bas y}{\ln\frac{k}{Q_{s0}}}}\ln \frac{Q_{s0}}{\Lambda}\right)\,\nonumber\\
&\times\, e^{2\sqrt{2\bas (Y-y) \ln \frac{kr}{2}}}e^{2\sqrt{2\bas y \ln \frac{k}{Q_{s0}}}}\label{sec-dep-1}
\end{align}

\item $k\gg 2/r\gg Q_s$. Repeating the by now familiar procedure yields
\beql{g-reg2}
g\approx 2\pi \int_{Q_s}^k dp p \frac{\partial \tilde N_G( p,  b, y)}{\partial\ln (1/p)}\,
\tilde f(\b r, \b k, Y-y)
\eeq
We observe that the cross section in this case is exactly the same as \eq{sec-dep-1}.

\item $Q_s\gg k\gg 2/r$:
\beq\label{g-reg3}
g\approx 2\pi \int_k^{Q_s} dp p \frac{\partial \tilde N_G( p,  b, y)}{\partial\ln (1/p)}\,
\tilde f(\b r, \b p, Y-y)
\eeq
With the help of \eq{sol-sat-ng} and \eq{f-dlog-2} we get
\beql{g-reg3-A}
g= 2 r S_0^2 \int_{-\infty}^{\infty} d\nu \frac{1}{1+2i\nu} \ e^{\frac{2\bar{\alpha}_s (Y-y)}{1+2i\nu}}\int_{k}^{Q_s} dp \ e^{-\ln^2 \frac{Q_s}{p}}\ln\frac{Q_s}{p}\left(\frac{rp}{2}\right)^{2i\nu}
\eeq
Now, using $\tau = \ln\frac{Q_s}{p}$ in place of $p$ 
\begin{align}\label{g-reg3-B}
&\int_k^{Q_s} dp\, e^{-\ln^2\frac{Q_s}{p}}\ln\frac{Q_s}{p} p^{2i\mu}=Q_s^{2i\mu+1}\int_0^{\ln(Q_s/k)} d\tau \,\tau\, e^{-\tau^2-\tau(1+2i\mu)}\nonumber\\
& \approx Q_s^{2i\mu+1}\int_0^\infty d\tau \,\tau\, e^{-\tau^2} =\frac{1}{2}Q_s^{2i\mu+1}\,.
\end{align}

Putting everything together yields
\begin{align}
\frac{d\sigma^{q\bar q+A}(\b k, y; \b r)}{d^2kdy}&=\frac{\bas C_F S_0^2}{\pi^{5/2} k^2} \int d^2b\, \frac{1}{(\ln\frac{r Q_s}{2})^{1/4}(2\bas (Y-y))^{1/4}}e^{2\sqrt{2\bas (Y-y) \ln \frac{rQ_s}{2}}}\label{sec-dep-3}
\end{align}

\item $Q_s\gg 2/r\gg k$:
\beql{g-reg4}
g\approx 2\pi \int_{2/r}^{Q_s} dp p \frac{\partial \tilde N_G( p,  b, y)}{\partial\ln (1/p)}\,
\tilde f(\b r, \b p, Y-y)
\eeq
This case is similar to the previous one except the the lower limit of the integral in \eq{g-reg3-A}, $k$,  is now replaced by $1/r$. However, for very large $Q_s$, the integral over $p$  is independent of the lower limit of integration as is clear from \eq{g-reg3-B}. We conclude thereby that the cross section in this case coincides with \eq{sec-dep-3}.

\item $2/r\gg k\gg Q_s$:
\beql{g-reg5-0}
g\approx 2\pi \int_{Q_s}^k dp p \frac{\partial \tilde N_G( p,  b, y)}{\partial\ln (1/p)}\tilde f(\b r, \b k, Y-y)+2\pi \int_k^{2/r} dp p \frac{\partial \tilde N_G( p,  b, y)}{\partial\ln (1/p)}\,
\tilde f(\b r, \b p, Y-y)
\eeq
The first of these integrals reads using \eq{g-reg1-A} and \eq{f-dlog-1}
\begin{align}
&2\pi\int_{Q_s}^k dp p \frac{\partial \tilde N_G( p,  b, y)}{\partial\ln (1/p)}\tilde f(\b r, \b k, Y-y)\nonumber\\
=& \frac{1}{32}\frac{(\ln\frac{k}{Q_{s0}})^{3/4}}{(2\bas y)^{5/4}}\frac{  1+\sqrt{\frac{2\bas y}{\ln\frac{k}{Q_{s0}}}}\ln \frac{Q_{s0}}{\Lambda}}{\left(2\bas (Y-y) \ln\frac{2}{rQ_{s0}} \right)^{1/4}}\,
Q_{s0}^2r^2\,
e^{2\sqrt{2\bas y \ln \frac{k}{Q_{s0}}}}\,
e^{2\sqrt{2\bas (Y-y) \ln \frac{2}{rQ_{s0}}}}
 \label{g-reg5-A0}
\end{align}
The second one is done by substituting \eq{N-lin-dif} and the integral form \eq{tilde-f} (it is useful to note that $\partial \tilde N_G/\partial \ln (1/p)\approx 2 \tilde N_G$) and then integrating over $p$ in the leading log approximation (i.e. treating $\log p$ as a constant) followed by the saddle point integral over $\nu$. We have
\begin{align}
&2\pi\int_k^{2/r}dp p \frac{\partial \tilde N_G( p,  b, y)}{\partial\ln (1/p)}\tilde f(\b r, \b p, Y-y)\nonumber\\
=&Q_{s0}^2r^2\frac{(\ln\frac{k}{Q_{s0}})^{1/4}(\ln\frac{2}{kr})^{1/4}\left( 1+\sqrt{\frac{2\bar{\alpha}_s y}{\ln\frac{k}{Q_{s0}}}}
\ln\frac{Q_{s0}}{\Lambda}    \right)}{2(2\bas y)^{3/4}(2\bas (Y-y))^{3/4}} 
e^{2\sqrt{2\bas y \ln \frac{k}{Q_{s0}}}}e^{2\sqrt{2\bas (Y-y) \ln \frac{2}{kr}}}
 \label{g-reg5-A1}
\end{align}
 Substitution of \eq{g-reg5-A0} and \eq{g-reg5-A1}  into \eq{xdip5} gives for the cross section 
\begin{align}\label{sec-dep-5}
\frac{d\sigma^{q\bar q+A}(\b k, y; \b r)}{d^2kdy}=\frac{\as C_F}{\pi^3 k^2}
\int d^2b\,
Q_{s0}^2r^2\frac{(\ln\frac{k}{Q_{s0}})^{1/4}(\ln\frac{2}{kr})^{1/4}\left( 1+\sqrt{\frac{2\bar{\alpha}_s y}{\ln\frac{k}{Q_{s0}}}}
\ln\frac{Q_{s0}}{\Lambda}    \right)}{2(2\bas y)^{3/4}(2\bas (Y-y))^{3/4}} 
&  \nonumber\\
\times
\,
e^{2\sqrt{2\bas y \ln \frac{k}{Q_{s0}}}}e^{2\sqrt{2\bas (Y-y) \ln \frac{2}{kr}}}\left[ 1+ \frac{(\ln\frac{k}{Q_{s0}})^{1/2}(2\bas (Y-y))^{1/2}}{ (2\bas y)^{1/2}(\ln \frac{2}{kr}\ln\frac{2}{rQ_{s0}})^{1/4} }\right]&
\end{align}

\item $2/r\gg Q_s\gg k$: 
\beql{g-reg6}
g\approx 2\pi \int_{Q_s}^{2/r} dp p \frac{\partial \tilde N_G( p,  b, y)}{\partial\ln (1/p)}\,
\tilde f(\b r, \b p, Y-y)
\eeq
Repeating the steps leading to \eq{g-reg5-A1} and noting \eq{ss}  we finally get 
\begin{align}\label{sec-dep-6}
\frac{d\sigma^{q\bar q+A}(\b k, y; \b r)}{d^2kdy}=\frac{\as C_F}{\pi^3 k^2}
\int d^2b\,Q_{s0}^2r^2\frac{(\ln\frac{2}{rQ_{s0}})^{1/4}
\ln\frac{Q_{s0}}{\Lambda} }{2^{5/2}(2\bas y)^{3/4}(2\bas (Y-y))^{3/4}} %&\nonumber \\
e^{4\sqrt{2}\bas y}e^{2\sqrt{2\bas (Y-y) \ln \frac{2}{Q_{s0}r}}}&
\end{align}
Eqs.~(\ref{sec-dep-1})--(\ref{sec-dep-6}) represent the dipole--nucleus inclusive cross section in all kinematic regions. 

\end{enumerate}

%%%%%%%
\section{Gluon production at the leading order in asymptotic regions}\label{sec:assym}
 
The DIS inclusive cross section is obtained from the dipole--nucleus one using \eq{xsect1}.  Integration over the dipole size $r$ and momentum fraction $z$ can be carried out for $Q\gg \Lambda,m$. In this case the largest contribution stems from the transversely polarized virtual photon.  Setting $m_f=0$ in \eq{w-f} we write \eq{xsect1} as 
 \begin{align}\label{xsect3}
\frac{d\sigma^{\gamma^*A}(\b k,y;Q)}{d^2kdy}= \frac{N_c}{\pi^2}\sum_f\frac{\alpha_\text{em}^f}{\pi}\int d^2 r \int_0^1 dz\, & Q^2z(1-z) \,K_1^2\left( r Q\sqrt{z(1-z)}\right)\, \nonumber\\
&\times\left[z^2+(1-z)^2\right]\frac{d\sigma^{q\bar q+A}(\b k, y; \b r)}{d^2kdy}\,.
\end{align}
At large $Q$ the dominant contribution to the $z$-integral arises from $z\to 0,1$. This corresponds to either quark or antiquark carrying most of the photon's energy.  These limits are symmetric, therefore we can calculate the $z$-integral for $z\to 0$ and multiply the result by 2.  Thus,
\begin{align}
\frac{d\sigma^{\gamma^*A}(\b k,y;Q)}{d^2kdy}&\approx \frac{N_c Q^2}{\pi^2} \frac{2\alpha_\text{em}}{3}\int_{4/Q^2}^\infty d r^2 \frac{d\sigma^{q\bar q+A}(\b k, y; \b r)}{d^2kdy}\,2\int_0^\infty dz\, z\, K_1^2\left( r Q\sqrt{z}\right)\nonumber\\
&=\frac{8N_c }{3\pi^2 Q^2}\frac{2\alpha_\text{em}}{3}\int_{4/Q^2}^\infty \frac{d r^2}{r^4}\frac{d\sigma^{q\bar q+A}(\b k, y; \b r)}{d^2kdy}\,,\label{xsect5}
\end{align}
where we took into account only three light quarks. To set the low limit of integration in \eq{xsect5} we noted that integrand in \eq{xsect3} peaks at  $rQ\sim 1/\sqrt{z(1-z)}\ge 2$. Upon substitution of \eq{xdip5} into \eq{xsect5} we get
\beql{xsect7}
\frac{d\sigma^{\gamma^*A}(\b k,y;Q)}{d^2kdy}=\frac{16\as\alpha_\text{em}}{9\pi^5}\frac{N_c C_F }{ Q^2k^2}
\int d^2b
\int_{4/Q^2}^\infty \frac{d r^2}{r^4}\int d^2p\, \frac{\partial \tilde N_G( p,  b, y)}{\partial\ln (1/p)}\,
\tilde f(\b r, \b p-\b k, Y-y)\,.
\eeq

To determine the cross section for gluon production in DIS it is convenient to do integral over $r$ before we integrate over $\nu$ in $\tilde f$. We thus define an auxiliary function 
\beql{h-fun}
d(Q,p,y)= \int_{\frac{4}{Q^2}}^\infty \frac{d r^2}{r^4} \tilde f(\b r, \b p, y)\,.
\eeq
Employing \eq{tilde-f} in \eq{h-fun} we obtain the Mellin representation of $d$
\begin{align}\label{h-melin}
d(Q, p,y)=\frac{Q}{2\pi p} \int_{-\infty}^{\infty} d\nu e^{2 \as\chi(\nu)y}\left(\frac{p}{Q}\right)^{2i\nu}\frac{\Gamma(\frac{1}{2}-i\nu)}{(\frac{1}{2}-i\nu)\Gamma(\frac{1}{2}+i\nu)(2i\nu+1)^2}
\end{align}
Inasmuch as we are interested only in asymptotic behavior of $d$, which we will derive using the saddle-point approximation, we can write in view of \eq{tilde-f}
\beql{f-h}
d(Q, p,y)=\frac{Q^2}{4}\tilde f(2/Q, \b p, y)\frac{1}{\frac{1}{2}-i\nu_\text{sp}}
\eeq
where $\nu_\text{sp}$ is a saddle point  given by one of the formulas \eq{sp2},\eq{spp},\eq{spp2}.  In particular, using \eq{f-dif}, \eq{f-dlog-1} and \eq{f-dlog-2} in \eq{f-h} yields
\begin{align}\label{Z-f-dif}
d(Q, \b p, y)&=\frac{Q}{4p}\frac{1}{\sqrt{14\pi\zeta(3)\bas\, y}}\, e^{(\alpha_P^{(0)}-1)y}\, e^{-\frac{\ln^2\frac{p }{Q}}{14\zeta(3)\bas\, y}}\,,\qquad  \as y\gg \ln^2\frac{p}{Q}\\
\label{Z-f-dlog-1}
d(Q, \b p, y)&=\frac{1}{4\sqrt{\pi}}\frac{(\ln\frac{Q}{Q_{s0}})^{1/2}}{\left(2\bas y)^{3/4}( \ln\frac{Q}{p} \right)^{1/4}} e^{2\sqrt{2\bas y \ln \frac{Q}{p}}}\,,\qquad Q\gg p\\
\label{Z-f-dlog-2}
d(Q, \b q, y)&=\frac{Q^2}{8\sqrt{\pi}p^2}\frac{1}{\left(2\bas y \ln\frac{p}{Q} \right)^{1/4}} e^{2\sqrt{2\bas y \ln \frac{p}{Q}}}\,,\qquad Q\ll p
\end{align}
Inspecting \eq{xsect7},\eq{h-fun},\eq{f-h},\eq{xdip5} and \eq{def-g} we get
\begin{align}\label{x-reg}
\frac{d\sigma^{\gamma^*A}(\b k,y;Q)}{d^2kdy}=\frac{4N_c \alpha_\text{em}\kappa}{9\pi^2}\frac{d\sigma^{q\bar q+A}(\b k,y;2/Q)}{d^2kdy}
\end{align}
where we denoted by $\kappa$ the logarithmic (or constant) factor $(1/2-i\nu_\text{sp})^{-1}$. 
Explicitly,
\beq\label{kappa}
\kappa = 2\left(\frac{\ln\frac{\max\{k,Q\}}{\min\{k,Q\}}}{2\bas (Y-y)}\right)^{1/2}\,,\,\,\text{if}\,\,\,  k,Q\gg Q_s\,;
\qquad 
\kappa =1\,,\,\, \text{if}\,\,\, k,Q\ll Q_s\,,
\eeq
Eq.~\eq{x-reg} together with the expressions of the inclusive dipole--nucleus cross section derived in \sec{sec:assym} provide the cross section for the inclusive gluon production in DIS at the leading logarithmic approximation.    

%%%%%%%%%%%
\section{NLO BFKL effects: energy  conservation }\label{sec:nlo}

\subsection{Dipole scattering amplitude}\label{sec:nlo.1}
As explained in the Introduction, one of the most important NLO effects is the momentum conservation.  BK equation modified to account for the energy  conservation reads \cite{Kormilitzin:2010at,Gotsman:2004xb} 
\begin{align}\label{BKcon}
\frac{\partial N(\b r,\b b, y)}{\partial y}= \frac{\bas}{2\pi}\left( 1-\frac{\partial }{\partial y}\right)
\int d^2r'\frac{\b r^2}{\b r'^2(\b r-\b r')^2}\left\{ N(\b r', \b b, y)+N(\b r-\b r', \b b, y)+N(\b r, \b b, y)\right.&
\nonumber\\
\left. -N(\b r', \b b, y)N(\b r-\b r', \b b, y)\right\}\,.&
\end{align}
In this section we discuss solution to this equation in dilute and saturation regimes. 

%%%%%%
\subsubsection{Dilute regime}

Consider first the dilute regime. It is advantageous to represent  $N$ as the double Mellin transform 
\begin{align}\label{exp-1}
N(\b r, \b b, y) = \int_{-i\infty}^{i\infty}\frac{d\omega}{2\pi i}\int_{-i\infty}^{i\infty}\frac{d\gamma}{2\pi i}\mathcal{ N}(\gamma,\b b, \omega)\frac{e^{\omega y +\gamma \xi - \xi}}{\omega - 2\bas\chi_1(\gamma,\omega) }\,,
\end{align}
where we introduced a new dimensionless variable $\xi= \ln(1/r^2Q_{s0}^2)$.    The anomalous dimension $\gamma$ is related to the Mellin variable $\nu$ that we have used so far  as $\gamma=1/2-i\nu$, so that the LO BFKL eigenvalue function is $\chi(\nu)= \chi(i(\gamma-1/2))$, see \eq{chi}.  $\chi_1(\gamma,\omega)$ denotes the NLO  BFKL eigenvalue function.  
In the dilute regime the $N^2$ term in the r.h.s.\ of \eq{BKcon} can be neglected. Substituting \eq{exp-1} into \eq{BKcon} one arrives at the following relation between the Mellin variables 
\begin{align}
\omega = 2\bas \chi_1(\gamma,\omega) = 2\bas (1-\omega) \chi\left(i(\gamma-1/2)\right)\,,
\end{align}
%%%%
\begin{figure}[ht]
\begin{tabular}{cc}
      \includegraphics[height=4.5cm]{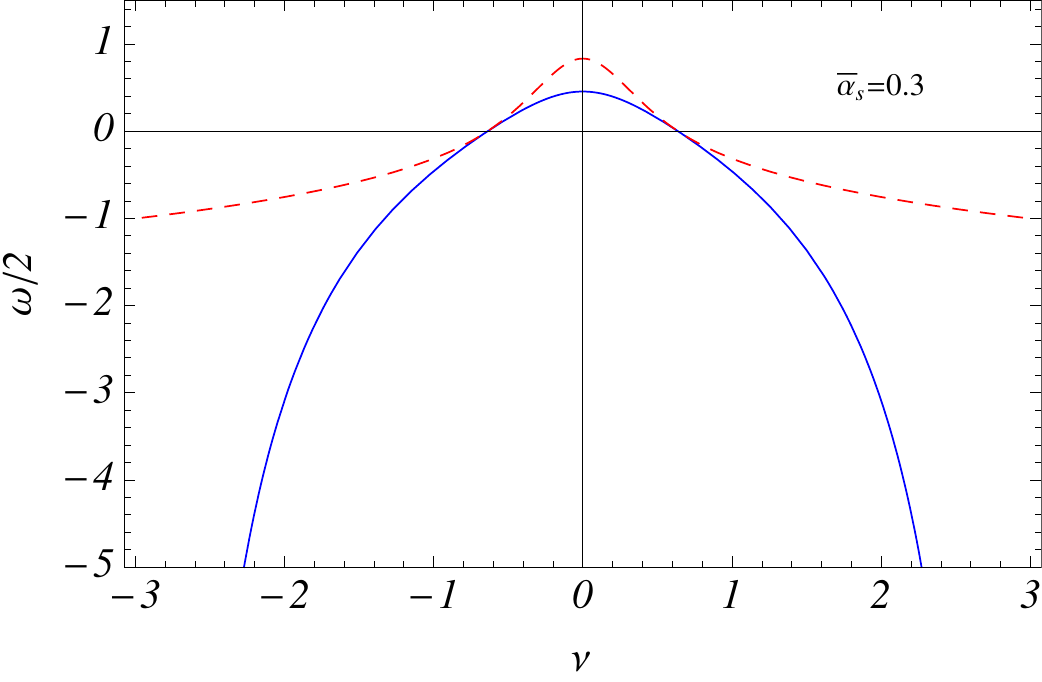} &
      \includegraphics[height=4.5cm]{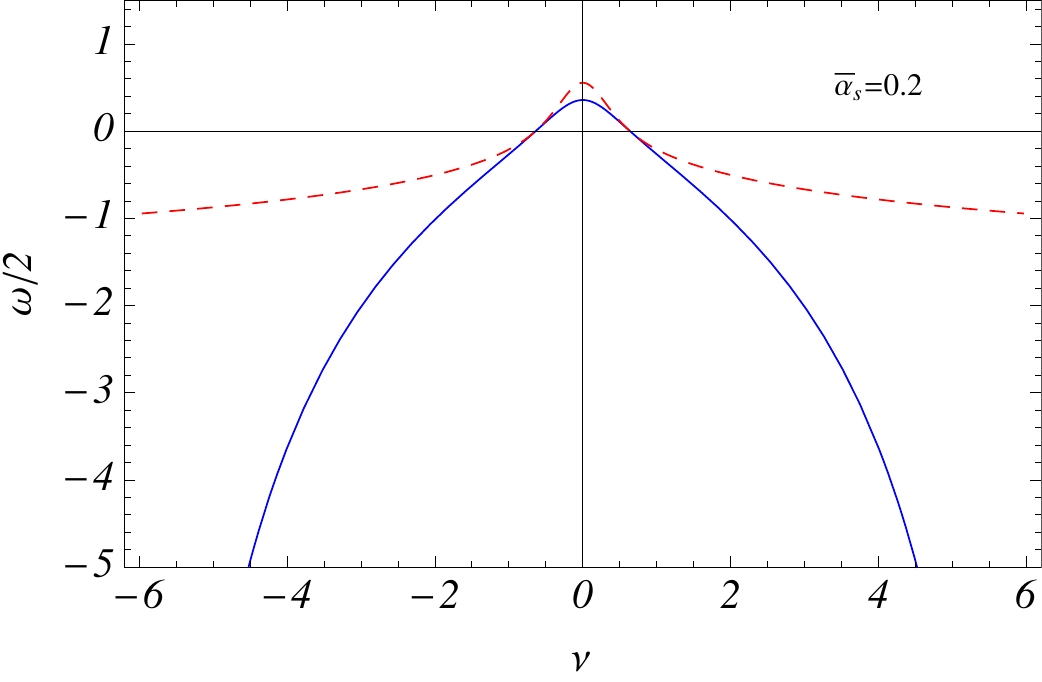}\\
      $(a)$ & $(b)$ 
      \end{tabular}
  \caption{$\omega(\nu)$ for (a) $\bas=0.3$ and (b) $\bas=0.2$. LO and NLO are represented by dashed (red) and solid (blue) lines respectively. Notice the different $\nu$ ranges of the two plots. }
\label{fig:omega}
\end{figure}
%%%%%
with the explicit solution for $\omega$ 
\begin{align}\label{wnlo}
\omega = \frac{2\bas \chi(\nu)}{1+2\bas \chi(\nu)}\,.
\end{align}
This solution is plotted in \fig{fig:omega}.  $\omega$ diverges at $\nu=\nu^*$ satisfying $2\bas \chi(\nu^*)=-1$. As $\bas\to 0$, $\omega$ approaches the LO expression while $\nu^*\to \pm\infty$.  At $\gamma\to 0$, i.e.\ $i\nu\to 1/2$, $\chi\approx 1/(1-2i\nu)=1/2\gamma$ and \eq{wnlo} yields
\beql{levin-mod}
\gamma(\omega)= \bas \left( \frac{1}{\omega}-1\right)\,.
\eeq
This can be used as a model of anomalous dimension that takes into account the  energy  conservation as suggested in \cite{Ellis:1993rb,AyalaFilho:1997du}. 
\footnote{Indeed, the anomalous dimension is proportional to the Mellin transform of the gluon splitting function
\beql{gtp}
\gamma(\omega)= \frac{\bas C_F}{\pi}
\int_0^1  P_\text{gg}(z)\, z^\omega dz \,.
\eeq
Energy conservation then implies that
\beql{gtp1}
 \gamma(1)=\frac{\bas C_F}{\pi}
\int_0^1  P_\text{gg}(z)\, z dz =0\,.
\eeq
}

Integrating \eq{exp-1} over $\omega$ we obtain
\begin{align}\label{n-2}
N(\b r, \b b, y) = \int_{-\infty}^{+\infty}d\nu \, C^A_\nu\, e^{\omega(\nu) y+\gamma\xi -\xi   } \,,
\end{align}
with $\omega(\nu)$ given by \eq{wnlo}. Remembering that in the dilute regime (and $N_c\gg 1$) $N_G=2N$, see \eq{Ng-N}, and using the same initial condition as in \eq{N-lin} we get
\beql{N-nlo}
N_G(\b r, \b b, y)= \int_{-\infty}^\infty d\nu\, \exp\left\{\frac{2\bas \chi(\nu)y}{1+2\bas \chi(\nu)}\right\}\, (rQ_{s0})^{1+2i\nu}\,\frac{1}{8\pi}\frac{1+(1-2i\nu)\ln\frac{Q_{s0}}{\Lambda}}{(1-2i\nu)^2}\,.
\eeq
This integral can be taken in the  double-logarithmic  approximation (DLA), which corresponds to keeping only one of the poles of $\chi$, namely $\chi(\nu)= 1/(1-2i\nu)$. Denote 
\beql{phase}
\phi(\xi,y)=\frac{2\bas \chi(\nu)y}{1+2\bas \chi(\nu)}-(1/2+i\nu)\xi\,.
\eeq
Then, in the DLA
\beql{phase-dla}
\phi(\xi,y)\approx \frac{\bas}{\gamma+\bas}y+\gamma \xi-\xi= 
2\sqrt{\bas y \xi}-\xi(1+\bas)+\frac{1}{2}\left(\gamma-\gamma_0\right)^2 \frac{2\xi^{3/2}}{(\bas y)^{1/2}}\,,
\eeq
where 
\beql{g-sp}
\gamma_0= \sqrt{\frac{\bas y}{\xi}} -\bas
\eeq
is the saddle point. Substituting \eq{phase-dla} into \eq{N-nlo} and integrating over the saddle point gives
\beql{N-nlo-dil}
N_G(\b r, \b b, y)=\frac{1+2\gamma_0\ln\frac{Q_{s0}}{\Lambda}}{32\pi^{1/2}\gamma_0^2}\frac{(\bas y)^{1/4}}{\ln^{3/4}\frac{1}{r^2Q_s^2}}(r^2Q_s^2)^{1+\bas} e^{2\sqrt{\bas y \ln\frac{1}{r^2Q_s^2}}}\,.
\eeq
The most important correction due to energy conservation requirement is steeper dependence of the scattering amplitude on $r$. 

%%%%
\subsubsection{Saturation momentum}
To determine the saturation momentum, we need to find a set of  lines in the $y,\xi$ plane along which the amplitude is constant. In the DLA approximation this is equivalent  to the requirement that the phase \eq{phase-dla} be constant, i.e.\ $ 2\sqrt{\bas y \xi}-\xi(1+\bas)=0$. Denoting solution to this equation as $\xi_s(y)$  we obtain
\beql{qs-dla}
Q_s^2=Q_{s0}^2e^{\xi_s}= Q_{s0}^2\,e^\frac{4\bas y}{(1+\bas)^2}\,.
\eeq
Energy dependence of the saturation momentum becomes more gradual compared to the LO. 

\begin{figure}[ht]
\begin{tabular}{cc}
      \includegraphics[height=4.6cm]{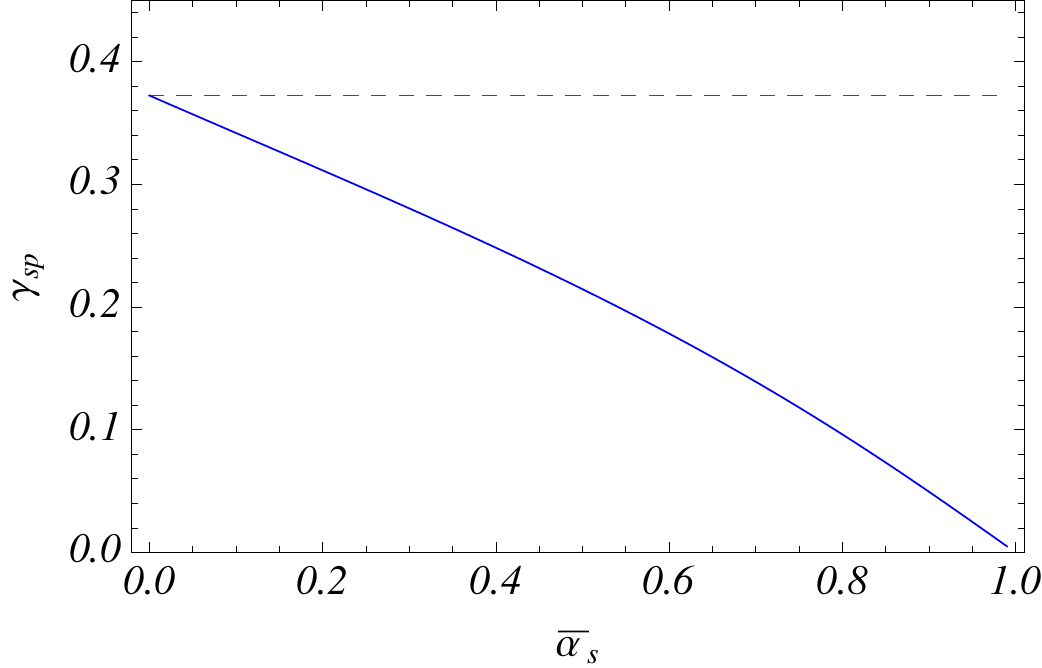} &
      \includegraphics[height=4.6cm]{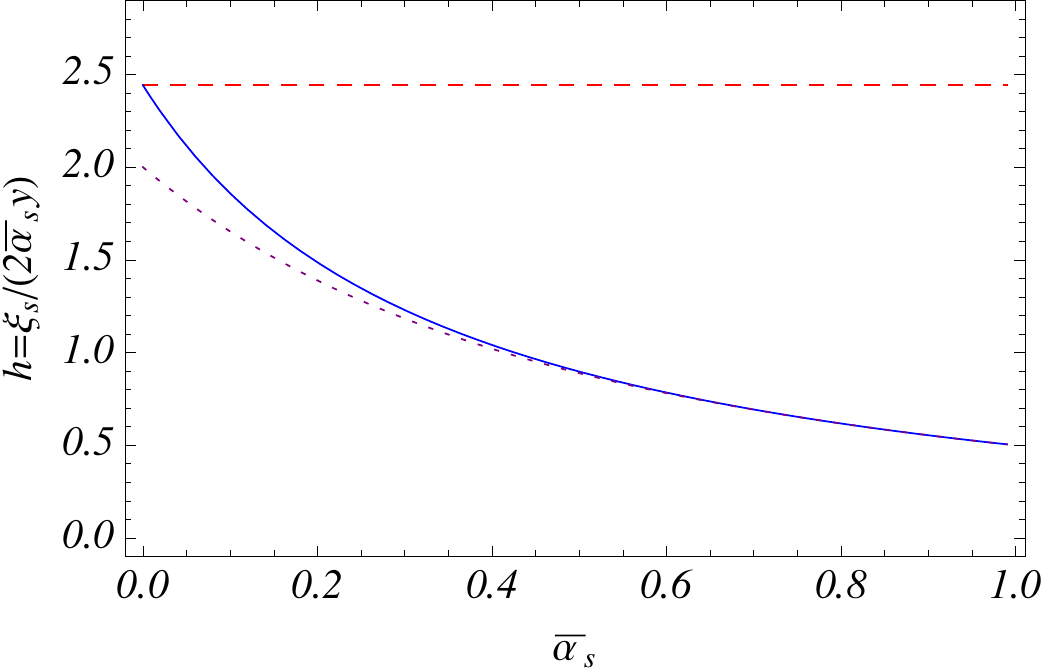}\\
      $(a)$ & $(b)$ 
      \end{tabular}
  \caption{ (a) Solution for the saddle point equation \eq{gsp} $\gamma_\text{sp}(\bas)$: solid blue line is NLO (energy conservation), dashed red line is LO. (b) Function $h(\bas)$ defined in \eq{gen-ss}: solid blue line is NLO, dotted (purple) is its DLA \eq{qs-dla} and  dashed  (red) is LO.}
\label{fig:ss-nlo}
\end{figure}

A more accurate evaluation of the saturation momentum requires solving the following two equations \cite{Mueller:2002zm}:
\begin{subequations}\label{sat-eq}
\begin{align}
&\phi=\frac{2\bas \chi(\gamma)y}{1+2\bas \chi(\gamma)}+\gamma\xi-\xi\,=0\\
&\frac{\partial\phi}{\partial\gamma}= \frac{2\bas \chi'(\gamma)y}{1+2\bas \chi(\gamma)}
-\frac{(2\bas)^2 \chi(\gamma)\chi'(\gamma)y}{(1+2\bas \chi(\gamma))^2}+\xi=0\,.
\end{align}
\end{subequations}
The first one determines the line on $y,\xi$ plane where the  amplitude is stationary, while the  second one fixes the trajectory of the steepest descend  \cite{Mueller:2002zm}. Eliminating $y$ and $\xi$ from these equations we end up with an equation for the saddle-point $\gamma_\text{sp}$:
\beql{gsp}
\chi'(\gamma_\text{sp})+\frac{1}{1-\gamma_\text{sp}}\chi(\gamma_\text{sp}) = \frac{2\bas\chi(\gamma_\text{sp})\chi'(\gamma_\text{sp})}{1+2\bas \chi(\gamma_\text{sp})}\,.
\eeq
 Employing \eq{chi} we write
\begin{align}
&\chi(\gamma) = \psi(1)-\frac{1}{2}\psi(\gamma)-\frac{1}{2}\psi(1-\gamma)\,,\label{gsp-a1}\\
& \chi'(\gamma)= -\frac{1}{2}\psi'(\gamma)+\frac{1}{2}\psi'(1-\gamma)\,.  \label{gsp-a2}
\end{align}
Saddle point in the LO is obtained as the solution to \eq{gsp} in the $\bas \to 0$ limit. Hence, dropping the r.h.s.\ of \eq{gsp} we obtain $\gamma_\text{sp}=0.37$. In the NLO 
 approximation $\gamma_\text{sp}$ depends on $\bas$ as shown in \fig{fig:ss-nlo}(a). As $\bas$ increases $\gamma_\text{sp}$ decreases and becomes closer to the experimental data. For a given  $\bas$ \eq{sat-eq} implies that
\beql{gen-ss}
Q_s^2= Q_{s0}^2  \exp\left\{ \frac{1}{1-\gamma_\text{sp}}\frac{2\bas\chi(\gamma_\text{sp})y}{1+ 2\bas\chi(\gamma_\text{sp})}\right\}\equiv  Q_{s0}^2e^{2\bas y \, h(\bas)}\,,
\eeq
Particularly, at the LO $h(\bas)= \frac{\chi(\gamma_\text{sp})}{1-\gamma_\text{sp}}=2.44$ independently of $\bas$. In \fig{fig:ss-nlo}(b) we show the NLO behavior of $h$ as given by \eq{gen-ss} and its DLA given by \eq{qs-dla}. Again we observe that the NLO correction makes the energy dependence of the saturation scale more gradual. This is understandable because the energy conservation reduces the phase space available for gluon emission.
%%%%
%%%%%

%%%%
\subsubsection{Saturation regime}
\begin{figure}[ht]
      \includegraphics[height=5cm]{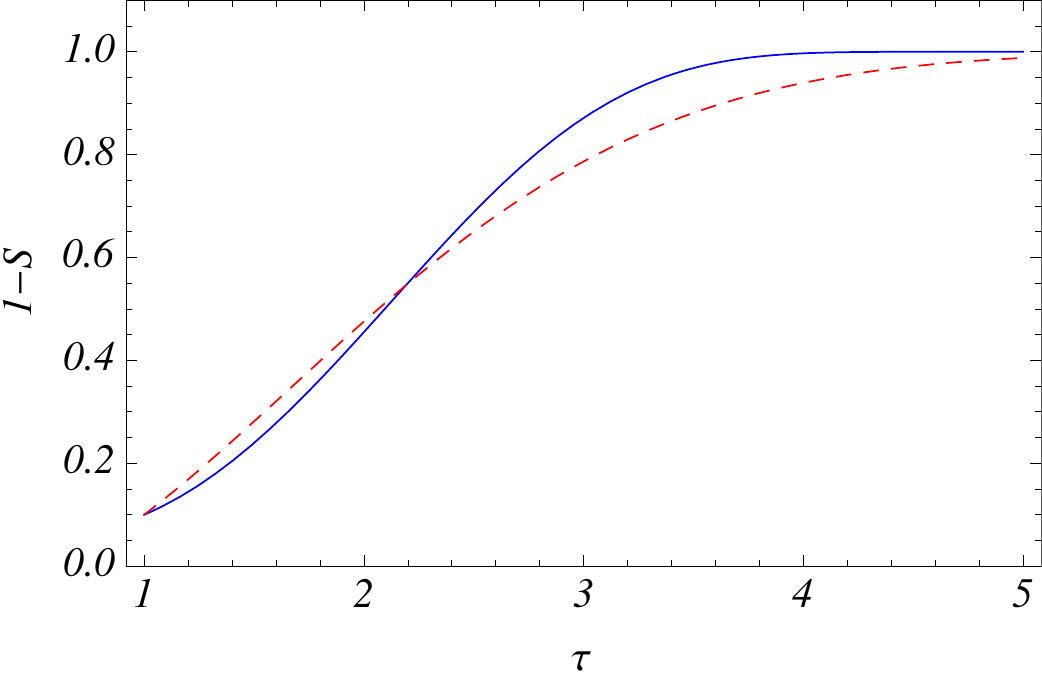} 
  \caption{Solution to the LO (dashed red line) and the modified (solid blue) BK  equations deeply in the saturation region $1<\tau <1/\bas$. The initial condition is   $S=0.9$  at $\tau=1$.    }
\label{fig:1-S}
\end{figure}

In the saturation region, \eq{BKcon} reads
\begin{align}\label{BKcon-sat}
\frac{\partial N(\b r,\b b, y)}{\partial y}= \bas\left( 1-\frac{\partial }{\partial y}\right)
\int_{2/Q_s^2}^{r^2} \frac{dr'^2}{r'^2}\left\{ N(\b r', \b b, y) -N(\b r', \b b, y)N(\b r, \b b, y)\right\}
\end{align}
We expect that the scattering amplitude will approach its unitarity limit as $y\to \infty$. Therefore, we are looking for a solution to \eq{BKcon-sat} in the form 
\beql{defS}
N=1-S
\eeq
where $S\ll 1$ is an element of the  scattering-matrix  of dipole $\b r$.  Now
\beql{eq-s}
-\frac{\partial S(r,y)}{\partial y }= \bas \left( 1-\frac{\partial }{\partial y}\right) \left\{ \ln(r^2Q_s^2)S(r,y)\right\}\,.
\eeq
We are interested in the scaling solution viz.\ we are looking for  a solution in the form $S(r,y)= S(\tau(r,y))$ where
\beql{tau}
\tau= \ln (r^2Q_s^2)= \ln(r^2Q_{s0}^2)+\frac{4\bas y}{(1+\bas)^2}\,,
\eeq
and we used \eq{qs-dla}. Introducing a new parameter that determines rapidity dependence of the saturation scale (in the DLA)
\beql{lambda}
\lambda = \frac{4\bas }{(1+\bas)^2}
\eeq
we write \eq{eq-s} as
\beql{eq-s-2}
\frac{\partial S}{\partial \tau}(\bas\lambda\tau-\lambda)=\bas(\tau-\lambda)S\,.
\eeq
It is easily integrated with the solution
\beql{s-sat-1}
S(\tau) = S_0e^\frac{\tau}{\lambda}(1-\bas\tau)^{\frac{1}{\bas\lambda}-1}\,,
\eeq
where $S_0$ is an integration constant that is determined by matching with the solution in the dilute regime. This is similar to the solution derived in \cite{Kormilitzin:2010at}. Note, that \eq{s-sat-1} is applicable only at $1<\tau\le 1/\bas$. 
Solution \eq{s-sat-1} is exhibited in \fig{fig:1-S}.
%%%%
%%%%%

%%%%%%%%%%%%%%%%%%%%%%%%%%%%%%
\subsection{Dipole density}

We proceed with the analysis of the NLO effects related to the energy  conservation in the dipole density. Using the result of the \sec{sec:nlo.1} we obtain in place of \eq{tilde-f}:
\begin{align}\label{f-sol-nlo}
&\tilde f(\b r, \b q, y)= \frac{r}{\pi q}\int_{-\nu^*}^{\nu^*} d\nu\, \exp\left\{\frac{2\bas \chi(\nu)y}{1+2\bas \chi(\nu)}\right\}\,\left(\frac{rq}{2}\right)^{2i \nu}\frac{\Gamma\left(\frac{1}{2}-i\nu\right)}{\Gamma\left(\frac{1}{2}+i\nu\right)(2i\nu+1)^2}\,,
\end{align}
where $\nu^*$ satisfy $1+2\bas \chi(\nu^*)=0$.
Similarly to our discussion in \sec{sec:ass-f}, we would like to find  asymptotic expressions for $\tilde f$ in various kinematic regions. Since the integrand in \eq{f-sol-nlo} is a steeply falling function of $\nu$ we can replace the limits of integration by $\nu^*=\pm\infty$.\footnote{Note, that we keep $\nu^*$ finite for the purpose of the numerical integration in \sec{sec:numer}.}
 \noindent
\begin{enumerate}

\item $\as y\gg \ln^2\frac{rq}{2}$. Expression in the exponent of \eq{f-sol-nlo} can be approximated as 
\begin{align}\label{e-d1}
\frac{2\bas \chi(\nu)y}{1+2\bas \chi(\nu)}\approx \frac{(\alpha_P^{(0)}-1)y}{\alpha_P^{(0)}}-\frac{14\zeta(3)\bas y}{[\alpha_P^{(0)}]^2}\nu^2\,.
\end{align}
We see that the pomeron intercept  became $\alpha_P^{(1)}= 2-1/\alpha_P^{(0)}$, while the ``diffusion constant" has increased by $1/ [\alpha_P^{(0)}]^2$, i.e.\ growth of $\tilde f$ with rapidity has slowed down, while diffusion has speeded up. The later observation has profound implications on diffractive gluon production (see \cite{Li:2008bm,Li:2008jz,Li:2008se} for in-depth discussion). For $\bas = 0.4$ the intercept is $\alpha_P^{(1)}= 1.5$ (compare with $\alpha_P^{(0)}=2.1$), which is in better agreement with the data. Eq.~\eq{f-dif} is modified as  follows
\beql{f-dif-nlo}
\tilde f(\b r, \b q, y)=\frac{r}{q}\frac{\alpha_P^{(0)}}{\sqrt{14\pi\zeta(3)\bas\, y}}\, e^{(\alpha_P^{(1)}-1)y}\, e^{-\frac{[\alpha_P^{(0)}]^2\ln^2\frac{rq}{2}}{14\zeta(3)\bas\, (Y-y)}}\,.
 \eeq

\item $rq<2$ and $\ln\frac{2}{rq}\gg \as y$. Expanding $\chi\approx 1/(1-2i\nu)$ we find the saddle point at 
\beql{sp-nlo}
2i\nu_1 = 1+2\bas -\sqrt{\frac{2\bas y}{\ln\frac{2}{rq}} }\,.
\eeq
Integration over the saddle-point and assuming  $\ln\frac{2}{rq}\ll y/\as$ yields
\beql{f-dlog-1-nlo}
\tilde f(\b r, \b q, y)=\frac{r^2}{8\sqrt{\pi}}\frac{(rq/2)^{2\bas}}{\left( 2\bas y \ln\frac{2}{rq} \right)^{1/4} 
\left[ 1-\sqrt{2\bas\frac{1}{y}\ln\frac{2}{rq}}  \right]} 
e^{2\sqrt{2\bas y \ln \frac{2}{rq}}}\,.
\eeq
 
\item $rq>2$ and $\as y\ll \ln\frac{rq}{2}\ll y/\as$. Now,  another pole in $\chi$ dominates $\chi\approx 1/(1+2i\nu)$ with the result
\beql{f-dlog-2-nlo}
\tilde f(\b r, \b q, y)=\frac{1}{2q^2\sqrt{\pi}}\frac{(2/rq)^{2\bas}}{\left(2\bas y \ln\frac{rq}{2} \right)^{1/4}
\left[ 1-\sqrt{2\bas\frac{1}{y}\ln\frac{rq}{2}}  \right]}
 e^{2\sqrt{2\bas y \ln \frac{rq}{2}}}\,.
\eeq

\end{enumerate}
Note, that in both cases  \eq{f-dlog-1-nlo} and \eq{f-dlog-2-nlo} the momentum dependence of the leading twist is modified by an additional power $2\bas$. This can have important consequences at high $Q^2$ and/or $k_T$. We are discussing this in more detail in \sec{sec:numer}.

%%%%%%%%%%%%%%%%%%%%%%%

%%%%%%%%%%
\section{Nuclear modification factor}\label{sec:nmf}

The nuclear modification factor is defined as 
\beql{nmf}
R_{\gamma^*A}= \frac{\int d^2b\,\frac{d\sigma_{\gamma^*A}}{d^2k\,dy\, d^2b}}
{A\,\int d^2b\,\frac{d\sigma_{\gamma^*p}}{d^2k\, dy\, d^2b}}\,.
\eeq
 In the logarithmic approximation \eq{x-reg} implies that the cross section for inclusive gluon production in DIS on a heavy nucleus is simply proportional to the cross section for inclusive gluon production 
by dipole of size $r= 2/Q$. Consequently, the nuclear modification factor \eq{nmf} can be approximated by 
\beql{appr-xs}
R_{\gamma^*A}\approx R_{q\bar q+A}\big|_{r=2/Q}\,.
\eeq
In the same approximation, $pA$ scattering can also be approximated as the $q\bar q+A$ one provided that we are interested in inclusive processes not too close in rapidity to the proton or nucleus  fragmentation region \cite{Li:2008se}.  
Atomic weight $A$ and rapidity $y$ dependence of incluisve cross section in $pA$ collisions at the leading logarithmic order was discussed in great detail in \cite{Kharzeev:2003wz} and we refer the interested reader to that paper. Here we will focus on \emph{the logarithmic derivative of the nuclear modification factor} defined as 
\beql{jdef} 
J=\frac{1}{R_{\gamma^*A}}\frac{\partial R_{\gamma^*A}}{\partial y}\,.
\eeq
Outside the saturation region  this observable is proportional to the difference between the anomalous dimension of the gluon distribution in the nucleus $\gamma^A$ and the one in the proton $\gamma^p$. If the coherence effects were negligible, the two anomalous dimensions would have been identical. This is not the case according to the theory of gluon saturation. As the result, the NMF is suppressed even at $k_T>Q_s$. Thus $J$  is especially sensitive probe of  the mechanism that leads to the suppression of the NMF for hadron production at small $x$.

Let us  relate $J$ to the difference of anomalous dimensions $\gamma^A-\gamma^p$. It follows from \eq{nmf} that 
\beql{za1}
J=\frac{\partial }{\partial y}\ln R_{\gamma^*A}= \frac{\partial }{\partial y}\ln \frac{d\sigma^{\gamma^*A}}{d^2k\, dy}-\frac{\partial }{\partial y}\ln \frac{d\sigma^{\gamma^*p}}{d^2k\, dy}\,.
\eeq
Using \eq{appr-xs} and \eq{xdip5},\eq{def-g} and assuming that the $b$-dependence factors out we derive
\beql{log-xs-1}
\frac{\partial }{\partial y}\ln R_{\gamma^*A}\approx \frac{\partial }{\partial y}\ln g^{A}\big|_{\b b=0}-\frac{\partial }{\partial y}\ln g^p\big|_{\b b=0}\,,
\eeq
where $g$ is the inclusive $q\bar q+A$ cross section modulo a constant factor, see \sec{sec:dipole}. We assigned superscripts $A$ and $p$ to  $g$ to indicate the two cases: $A>1$ and $A=1$ respectively. In the following we will omit the specification that $g$ is taken at zero impact parameter. Outside the saturation region we can employ the Mellin representation for $ N_G$ \eq{N-lin} and $\tilde f$ \eq{tilde-f}, substitute them into \eq{def-g}, take the LLA limit and obtain up to a pre-exponential factor
\beql{log-xs-2}
g^A\propto Q_0 r\, \exp\left[2\bas \chi(\nu_0)(Y-y)+2i\nu_0\ln \frac{rp}{2}
+2\bas \chi(\mu_0^A)y+2i\mu_0^A\ln \frac{Q_{s0}}{p}
\right]
\eeq
and analogously for $g^p$. Here $\nu_0$, $\mu_0^A$ are the saddle points in the Mellin transform of $\tilde f$ and $\tilde N_G$ respectively. The omitted pre-factor in \eq{log-xs-2} depends on momenta only logarithmically. Momentum $p$ stands for either $Q$ or $k$ depending on the kinematic region of interest. It is straightforward to verify that $g^A$ and $g^p$ obey the equations
\begin{align}\label{log-xs-3}
\frac{\partial g^A}{\partial y}= 2\bas [ \chi(\mu_0^A)-\chi(\nu_0)] g^A\,,\qquad \frac{\partial g^p}{\partial y}= 2\bas [ \chi(\mu_0^p)-\chi(\nu_0)] g^p\,.
\end{align}
This is just the Mellin transform of the BFKL equation. Plugging \eq{log-xs-3} into \eq{log-xs-1} we derive
\beql{log-xs-4}
J=2\bas \left[\chi(\mu_0^A)- \chi(\mu_0^p)\right]\approx 2\bas \chi'(\gamma_0^p)\, \left(\gamma_0^A- \gamma_0^p\right)\,.
\eeq
$\chi'(\gamma)$ is given by \eq{gsp-a2} and the saddle point $\gamma_0^p$ satisfies \eq{gsp}. 

Consider a few examples. Denote $p=\max\{ k_T,Q\}$. In the region $\ln \frac{p}{Q_{s0}}\gg \bas y$ we have (see e.g.\ \eq{spp}  and \eq{N-lin-dif})
\beql{n1}
\chi\approx \frac{1}{1-2i\mu}= \frac{1}{2\gamma}
\eeq
with the saddle point 
\beql{n2}
\gamma^A= \frac{1}{2}(1-2i\mu_0^A)= \frac{1}{2}\sqrt{\frac{2\bas y}   {\ln\frac{p}{\Lambda}+\ln\frac{\Lambda}{Q_{s0}}}}\approx 
\frac{1}{2}\sqrt{\frac{2\bas y}{\ln\frac{p}{\Lambda}}}\left( 1+\frac{\ln \frac{Q_{s0}}{\Lambda}}{2\ln\frac{p}{\Lambda}}\right)
\eeq
$\gamma^p$ is obtained by setting $Q_{s0}=\Lambda$. We see that in this kinematic region $\gamma^p< \gamma^A$. By dint of \eq{n1}  $\chi'(\gamma)<0$ implying that $J<0$. More precisely,
\beql{n3}
J= -\bas \frac{\ln\frac{Q_{s0}}{\Lambda}}{\sqrt{2\bas y \ln \frac{p}{\Lambda}}}\,.
\eeq
In the saturation region $\ln \frac{p}{Q_{s0}}\ll \bas y$, $\gamma^A$ effectively tends to zero as the dipole scattering amplitude saturates at  unity. Therefore,  in that region $\gamma^A<\gamma^p$, while $\chi\approx \frac{1}{2(1-\gamma)}$. Hence $\chi'(\gamma)>0$ implying that again $J<0$.  
Finally, in the diffusion region $\chi\approx 2\ln 2-7\zeta(3)\nu^2$ and we similarly obtain 
\beql{n4}
J= -\frac{\ln\frac{p}{\Lambda}\ln \frac{Q_{s0}}{\Lambda}}{7\zeta(3)\bas y^2}\,.
\eeq
Negativity of $J$ in all kinematic regions signifies the decrease of the inclusive cross section as a function of rapidity. The rate of the decrease depends on the absolute value of $J$.

%%%%%%%%%%%
\section{Numerical analysis}\label{sec:numer}

The numerical calculation of the inclusive hadron production is performed using Eqs.~\eq{xsect1},\eq{w-f},\eq{xdip3},\eq{dist-def}. We employed the bGCG model \cite{Kowalski:2006hc} for the forward dipole--nucleus scattering amplitude. The bCGC model is  reviewed in Appendix. Function $\tilde f$  is calculated using formula \eq{f-sol-nlo}. The gluon spectrum is then convoluted with the LO pion fragmentation function $\mathcal{F}_G$ as follows
\beql{frag}
\frac{d\sigma_{\pi}}{d^2k\,dy} =
\int_{z_\text{min}}^1 \frac{dz}{z^2} \, \frac{d\sigma_G}{d^2k\,dy}(k/z) \,
\mathcal{F}_G (z, k)\,.
\eeq
The fragmentation function is given in \cite{Kniehl:2000hk}.
   The total rapidity interval is taken to be $Y=10$, which is equivalent  to $x=e^{-Y}=4.5\cdot 10^{-5}$. The range of photon virtualities that we consider is $Q^2=2-37$~GeV$^2$. This kinematic region can be probed at the proposed Large Hadron electron Collider and its low $Q^2$ part at the Electron Ion Collider \cite{Boer:2011fh}.  The rapidity interval $y$  from the nucleus to the produced gluon is related to $x_{\Pom}$, a variable used in differctive DIS,   as $x_{\Pom}=e^{-y}$. We consider $y$ in a narrow  interval   $5\le y\le 7$ allowed by our formalism.  At larger $x$ and/or $x_\Pom$ the validity of the leading logarithmic approximation that we employ becomes uncertain.

\begin{figure}[t]
\begin{tabular}{cc}
      \includegraphics[height=4.6cm]{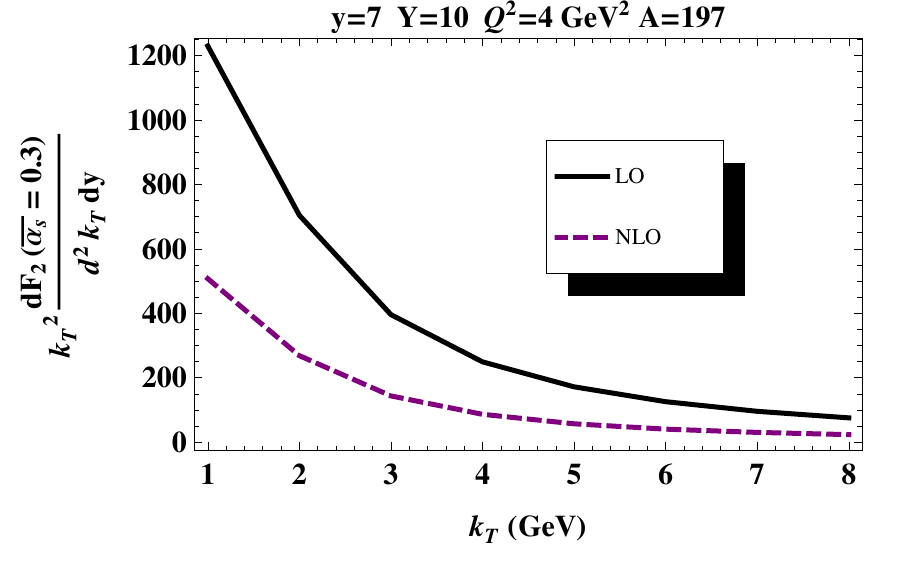} &
      \includegraphics[height=4.6cm]{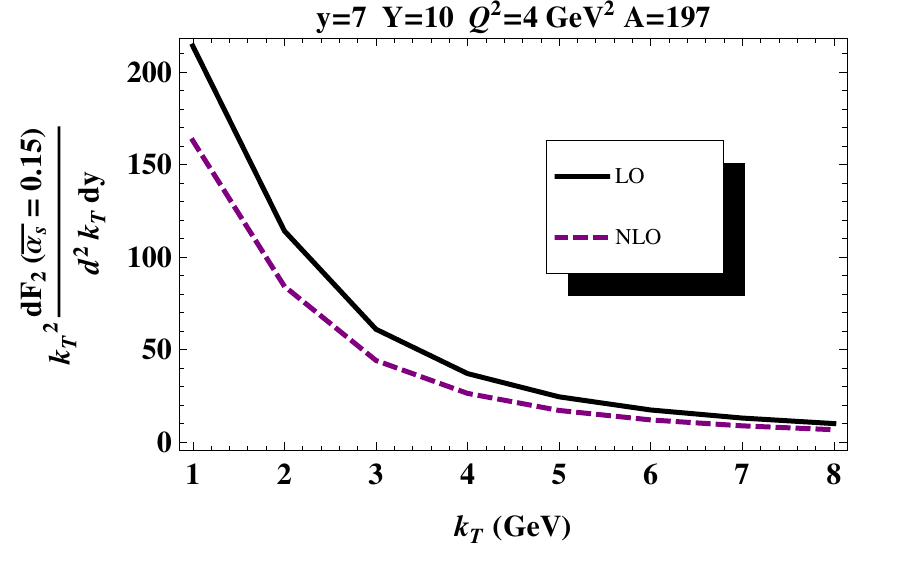}\\ [-2mm]
      $(a)$ & $(b)$ 
      \end{tabular}
  \caption{Comparison between the LO and NLO calculations of $k^2\frac{dF_2(x,Q^2;y)}{d^2k dy}$ as a function of $k_T$ at two values of coupling (a) $\bas=0.3$ and  (b) $\bas = 0.15$.}
\label{fig:F2a}
\end{figure}
\begin{figure}[t]
\begin{tabular}{cc}
      \includegraphics[height=4.6cm]{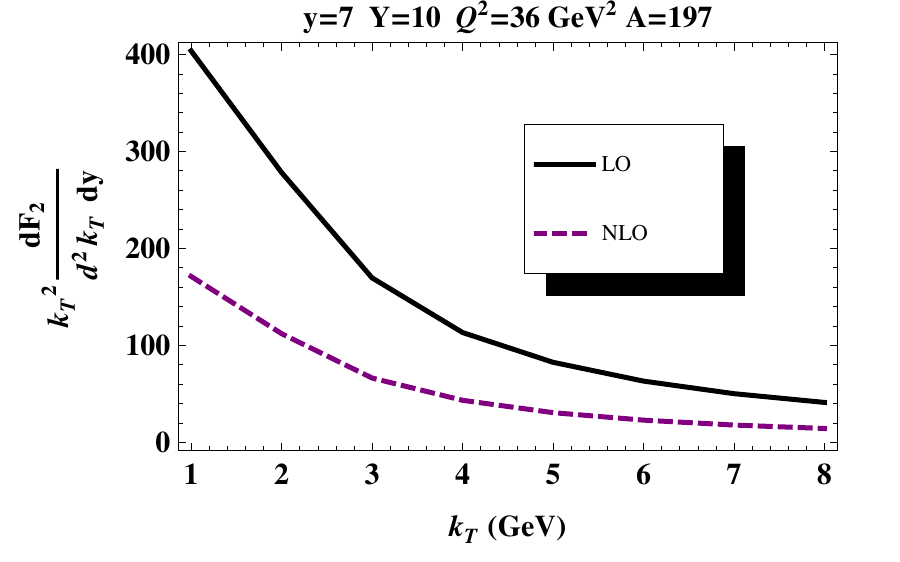} &
      \includegraphics[height=4.6cm]{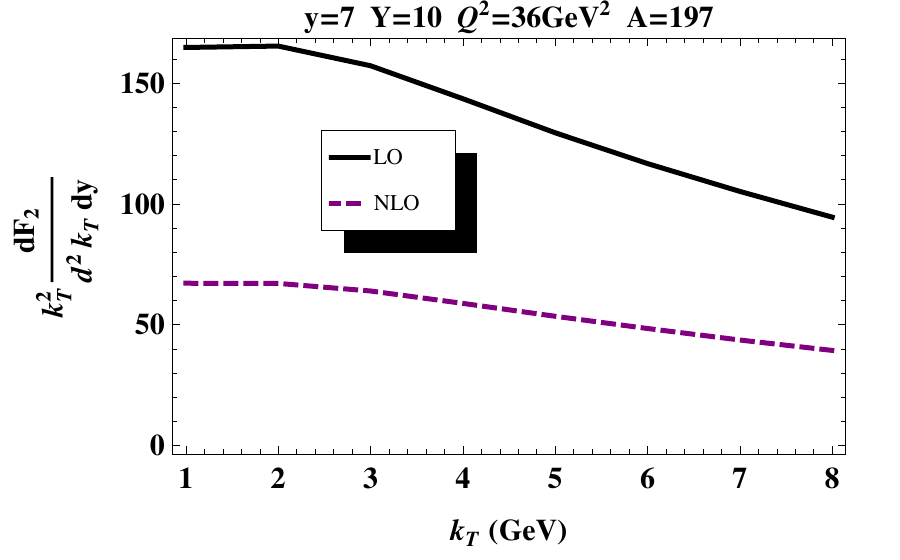} 
      \\ [-2mm]
      $(a)$ & $(b)$ \\[2mm]
      \end{tabular}
  \caption{Inclusive spectrum $k^2\frac{dF_2(x,Q^2;y)}{d^2k dy}$ of (a) pions, (b) gluons as a function of   $k_T$.}
\label{fig:F2}
\end{figure}
\begin{figure}[t]
\begin{tabular}{cc}
      \includegraphics[height=4.6cm]{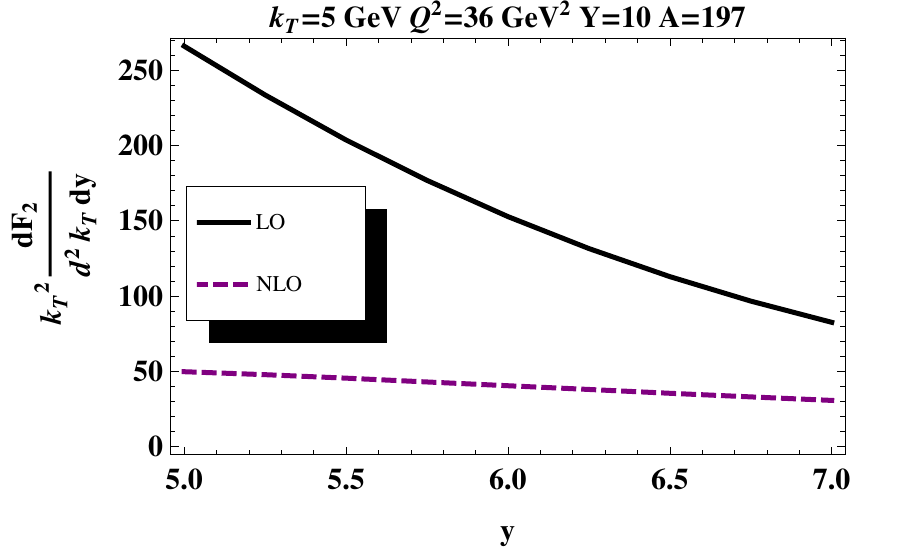} &
      \includegraphics[height=4.6cm]{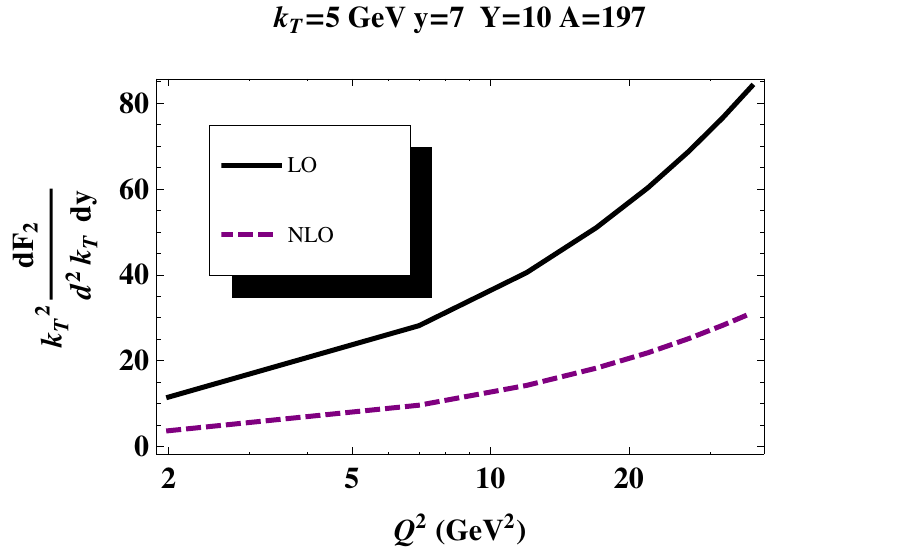}\\ [-2mm]
      $(a)$ & $(b)$ \\[2mm]
      \end{tabular}
  \caption{Inclusive hadron spectrum $k^2\frac{dF_2(x,Q^2;y)}{d^2k dy}$ as a function of  (a) $y$, (b) $Q^2$.}
\label{fig:F2c}
\end{figure}
\begin{figure}[t]
\begin{tabular}{cc}
      \includegraphics[height=4.6cm]{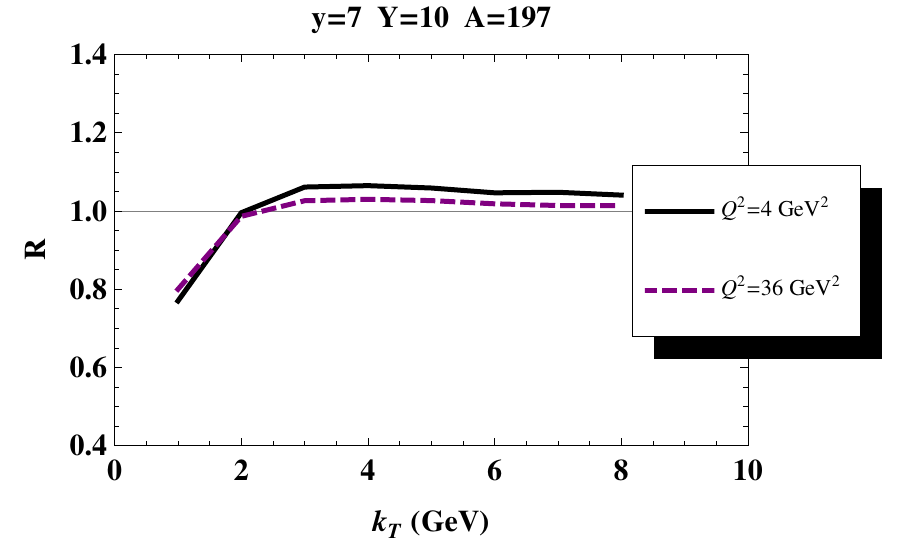} &
      \includegraphics[height=4.6cm]{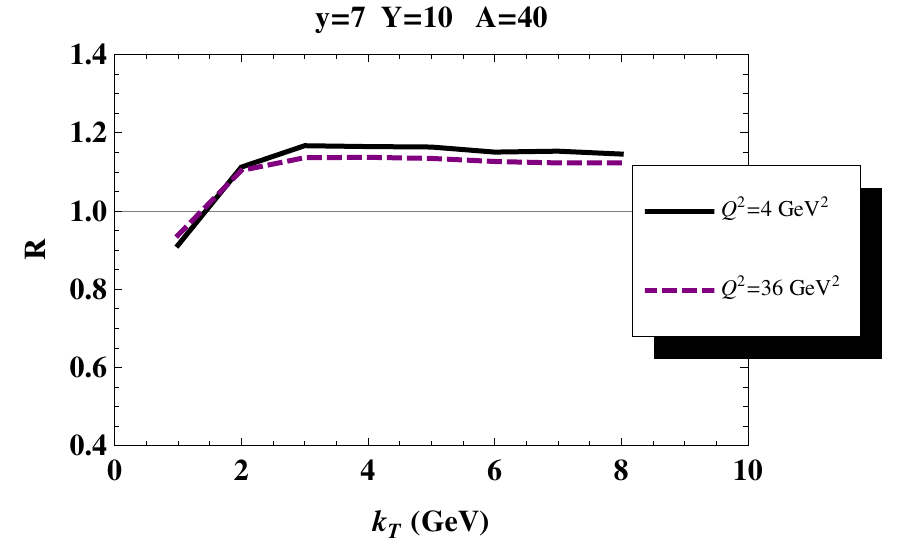}\\
      $(a)$ & $(b)$ \\
      \includegraphics[height=4.6cm]{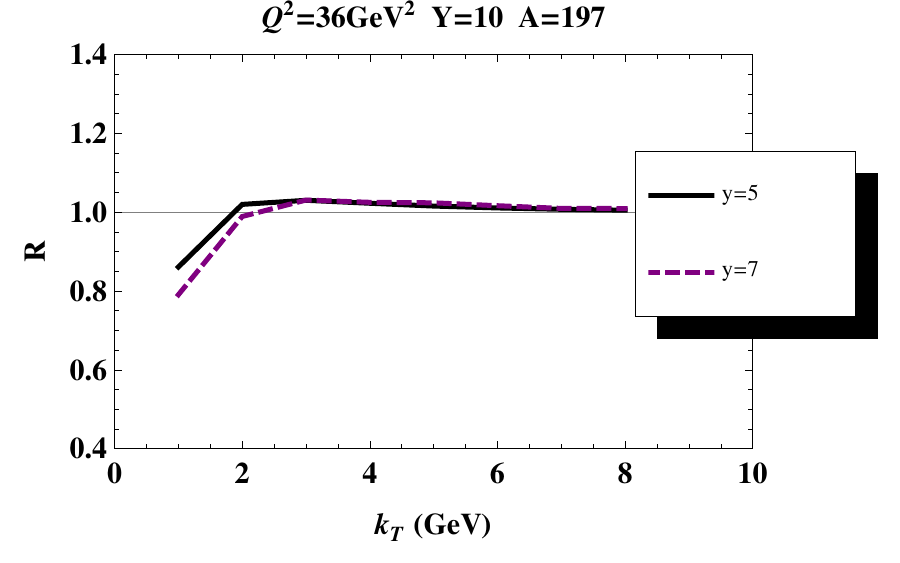}&
      \includegraphics[height=4.6cm]{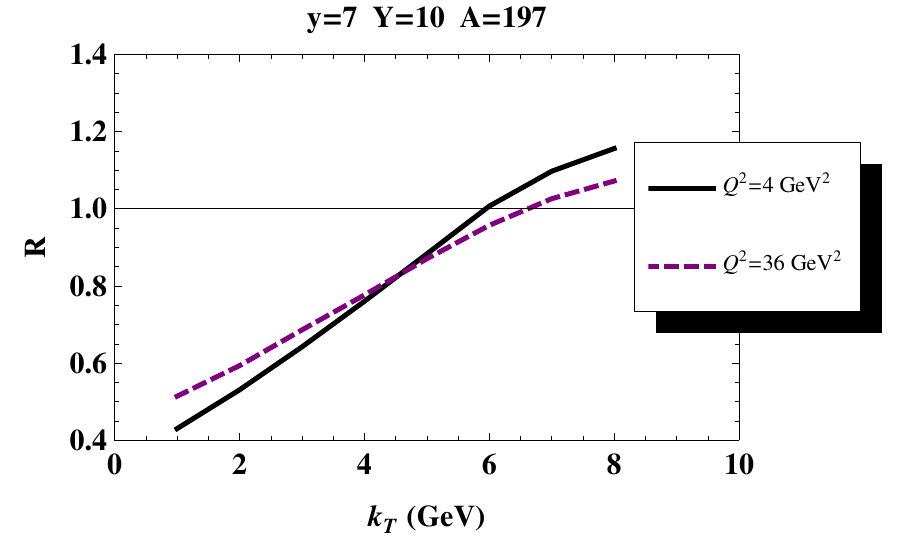}\\ [-2mm]
      $(c)$ & $(d)$
      \end{tabular}
  \caption{  Nuclear Modification Factor as a function of $k_T$ for (a)-(c) hadrons at various $A$, $y$ and $Q^2$; (d) gluons. All calculations include the NLO effects.  }
\label{fig:NMF}
\end{figure}
\begin{figure}[t]
\begin{tabular}{cc}
 \includegraphics[height=4.6cm]{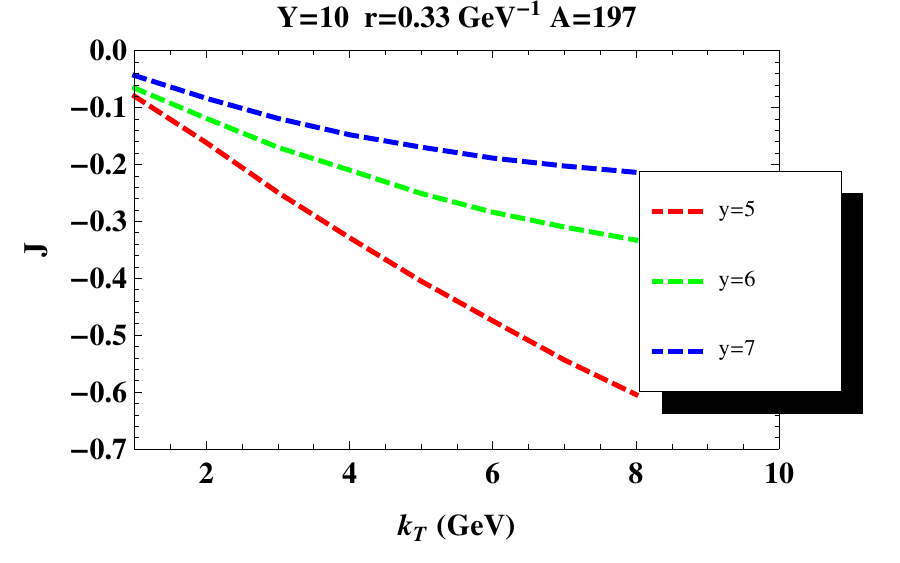} &
      \includegraphics[height=4.6cm]{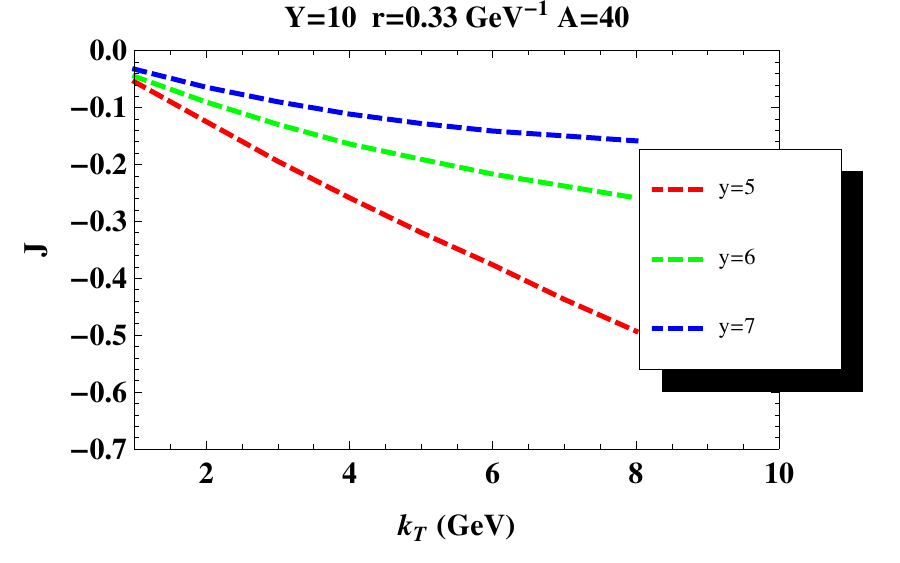}\\[-2mm]
   $(a)$ & $(b)$ \\[2mm]
      \includegraphics[height=4.6cm]{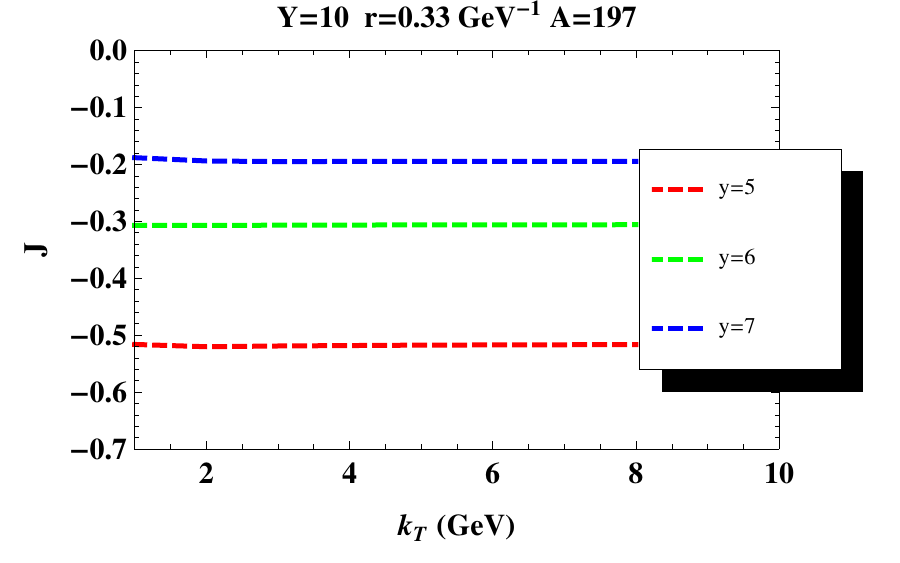} &
      \includegraphics[height=4.6cm]{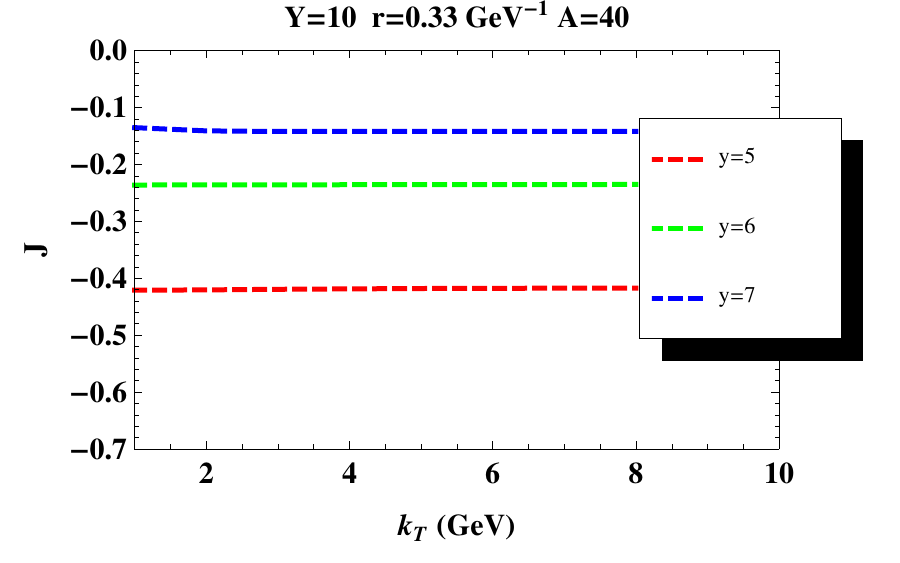}\\ [-2mm]
      $(c)$ & $(d)$ \\[2mm]
      \end{tabular}
  \caption{ Logarithmic derivative of NMF for dipole-nuleus scattering as a function for $k_T$ for (a),(b) gluons, (c),(d) hadrons. dipole size $r$, total rapidity $Y$ and nuclear wight $A$ are indicated on each plot. All calculations include the NLO effects.  }
\label{fig:Jqqbar}
\end{figure}

The results of our calculations are shown in Figs.(\ref{fig:F2a})--(\ref{fig:Jqqbar}). The NLO calculation shown in the figures refers  to the part of the NLO terms that are responsible for energy conservation. In \fig{fig:F2a},\ref{fig:F2} we plot the inclusive cross section normalized in the same way as the structure function  
\beql{F2}
\frac{dF_2(x,Q^2;y)}{d^2k dy}=\frac{1}{\alpha_{em}}\frac{Q^2}{4\pi^2}\frac{d\sigma^{\gamma*A}(x,Q^2;y)}{d^2k dy}\,.
\eeq
We observe that inclusive gluon production at NLO is suppressed compared with the LO case. This is because the anomalous dimension of dipole density at  NLO is smaller compared with that of  LO, as can be seen in  \fig{fig:ss-nlo}. This is expected since energy  conservation constrains the phase space available for hadron production. 
In \fig{fig:F2a} we demonstrate that the difference between the LO and NLO calculation is smaller at smaller values of coupling.

We see in \fig{fig:F2}(b) that at small $k_T$, the gluon production cross section follows $1/k_T^2$ behavior. Indeed, $1/k_T^2$ comes from the Lipatov vertex, whereas the gluon distribution in the nucleus is saturated and hence depends on momentum $k_T$ only logarithmically. This is  seen in  \eq{xdip3} where at small $k_T$ the integral tends to a constant leaving the $1/k_T^2$ pre-factor in front. Modification of the gluon spectrum due to fragmentation can be inferred by comparing  \fig{fig:F2}(a) and (b). 

The cross section grows with $Q^2$ and $x_{\Pom}$ logarithmically; both dependences are much steeper at the LO than in the NLO. We also note that energy conservation correction substantially reduces the cross section. However, the functional form of the $k_T$-spectrum does not change in the kinematic region that we studied, as we checked explicitly. We attribute this to that fact that the dominant contribution to the Mellin transform stems from anomalous dimension $\gamma\approx 1/2$ in both cases. We expect that at much larger $Q$ and $k_T$ the NLO $k_T$-spectrum becomes steeper than those in LO due to  additional factors $1/Q^{2\bas}$ or $1/k_T^{2\bas}$. However, assumptions of our model restrict our calculation only to the semi-hard values  of transverse momenta. 

The largest uncertainty in our numerical calculation of hadron spectrum comes from the oversimplified treatment of nuclei geometry. Instead of integrating with a realistic nuclear thickness $T(b)$ we  approximated the nuclear density by the step-function. Based on our previous experience with this type of numerical calculations we expect that a more accurate treatment of the nuclear density will only affect the overall normalization of the cross section.  From this perspective the ratios of the inclusive spectra should not be much affected by this uncertainty.

Our calculation of the Nuclear Modification Factor (NMF) as 
a function of $k_T$ for Au ($A=197$) and Ca ($A=40$) is displayed in \fig{fig:NMF}. The general feature of NMF is suppression at low $k_T$ and enhancement at larger $k_T$ (the later is often referred to as the Cronin effect). This is in contrast with the hadron production in $pA$ scattering where the Cronin effect gives way to the suppression at all $k_T$'s provided that the hadron rapidity $y$ is large enough. The reason for this difference is that whereas $pA$ scattering can be approximated by dipole-nucleus scattering  \cite{Li:2008se}, $\gamma^*A$ interaction is a superposition of many dipole-nucleus scatterings with different dipole sizes $r$, see \eq{xdip1}. At small $k_T$ NMF for dipoles of all sizes is suppressed \cite{Kharzeev:2003wz} and therefore we observe suppression of the resulting $R$ for DIS. On the other hand, the fact that $R>1$ at large $k_T$ implies that the inclusive cross section in that region is dominated by  dipoles whose individual scattering on the nucleus exhibits Cronin enhancement, i.e.\ they are not much effected by the low-$x$ evolution. Presence of such dipoles is ensured by evolution of the dipole density $n$, which happens if $Y-y\gg 1$.  Comparing Figs.\ref{fig:NMF}~(a)-(c) with (d) we note that due to fragmentation, NMF of hadrons is much slower function of $Q^2$, $y$ and $k_T$ than NMF of gluons. Additonally, fragmentation shifts the value of the transverse momentum at which NMF crosses unity  towards lower $k_T$.

%%%%
%%%%%
Another feature seen in \fig{fig:NMF} (especially (d)) is that suppression of NMF at low $k_T$ and its enhancement at high $k_T$ increases with the photon virtuality $Q^2$.
To understand the $Q^2$ dependence of the NMF we note that a typical term in its  twist expansion looks like 
\beql{twists1}
R\sim \left(\frac{1}{Q^2}\right)^{n(\gamma_A-\gamma_p)}\,,
\eeq
where $n\ge 1$  is an integer number. It implies that 
\beql{twists2}
\frac{\partial R}{\partial\ln Q^2}\approx -n(\gamma_A-\gamma_p)R.
\eeq
At large $k_T$ $\gamma^A>\gamma^p$ thus $ \frac{\partial R}{\partial\ln Q^2}<0$, whereas at small $k_T$ $\gamma^A<\gamma^p$ thus $ \frac{\partial R}{\partial\ln Q^2}>0$. This is indeed what we observe in \fig{fig:NMF}. Dependence of NMF on $y$ can be explained similarly.

%%%%%%%%
%%%%%
\fig{fig:Jqqbar} displays the logarithmic derivative of the NMF $J$ defined in \eq{jdef}. As we argued in \sec{sec:nmf} this quantity is proportional to the difference between the anomalous dimensions of the gluon distribution function in nucleus and proton, see\eq{log-xs-4}. Our analysis in \eq{n3},\eq{n4} indicates that $J$ is negative and decreases as the hadron rapidity $y$ increases, which is indeed seen in  \fig{fig:Jqqbar}. 
Similar trend  has been noticed in $pA$ collisions in \cite{Tuchin:2007pf}. We can also see the effect of fragmentation on $J$ by comparing \fig{fig:Jqqbar}(a),(b) with (c),(d). It is interesting that fragmentation completely erases the $k_T$ dependence, while leaving the $y$ dependence qualitatively similar. We think that experimental investigation of $J$ is of great interest as it emphasizes the difference between the (linear) gluon evolution in a heavy nucleus and in proton.

%%%%%%%%%%%%%%%%%%%%%%%%%%%%%%%%%%%%%%
\section{Summary}\label{sec:summary}

In this paper we studied the inclusive hadron production in DIS scattering at small $x$
using  the dipole model \cite{Mueller:1989st}. We presented the analytical formulas for the cross section in various kinematic regions and discussed the role of the energy conservation, which is perhaps the most important NLO correction. Employing the modified BK equation suggested in \cite{Kormilitzin:2010at,Gotsman:2004xb}, we derived the corresponding correction to the pomeron intercept and found that it is numerically closer to the phenomenological value than the LO result. We also computed the high energy asymptotic of the forward dipole-nucleus scattering amplitude. 

Motivated by possible low $x$ DIS experiments with heavy nuclei \cite{Boer:2011fh} we performed numerical calculations of the DIS inclusive cross section using the bCGC model \cite{Kowalski:2006hc}. The results are shown in Figs.~\ref{fig:F2a}--\ref{fig:Jqqbar}.
We noticed that the NLO effects generally tend to reduce the cross section and make it weaker function of its arguments as compared to the LO result. The nuclear modification factor exhibits suppression at low $k_T$ and enhancement at higher $k_T$ even at the largest hadron rapidities that we can address in our approach. To understand dependence of the NMF on rapidity better we introduced the logarithmic derivative of NMF $J$ and showed that it is proportional to the difference between the anomalous dimension of the gluon distribution function in nucleus and proton. Since this difference is non-vanishing only due to coherence effects, $J$ provides a direct measure of the effect of coherence on inclusive cross section. Figs.~\ref{fig:NMF},\ref{fig:Jqqbar} show dependence of NMF and $J$ on the photon virtuality $Q^2$, $x$ and  hadron rapidity $y$. We believe that our results may be helpful for experimental investigation of the low-$x$ regime of QCD in DIS.

%%%%%%%%%%%%%%%%%%%%%%%%%%%%%%%%
\acknowledgments
%I  am grateful to ... for many fruitful discussions of related problems. 
This work  was supported in part by the U.S. Department of Energy under Grant No.\ DE-FG02-87ER40371.

%%%%%%%%%%%%%%%%
\appendix
\section{bCGC model}\label{appA}

We performed  the numerical calculations  using  the bCGC model of the forward  dipole scattering amplitude \cite{Kowalski:2006hc}.  We treat the nuclei and proton profiles as step-functions; the saturation scales are assumed to scale with $A$ as $Q_s^2\propto A^{1/3}$. The advantage of this model -- besides its compliance with the known analytical approximations to the BK equation \cite{Iancu:2002tr} -- is that its parameters are fitted to the low $x$ DIS data. The explicit form of the scattering amplitude $N$   is given by
\beql{kmw}
N(\b r, 0, y)= \,\left\{
\begin{array}{cc}
\mathcal{N}_0\left( \frac{r^2\Q_s^2}{4}\right)^\gamma\,,&\quad r\Q_s\le 2;\\
1-\exp[-a\ln^2(br\Q_s)]\,,&\quad r\Q_s\ge 2\,,
\end{array}\right.
\eeq
where $\Q_s^2$ is the the \emph{quark} saturation scale related to the \emph{gluon} saturation scale $Q_s^2$ -- which we have called simply the `saturation scale' throughout the paper -- by $\Q_s^2= (4/9)Q_s^2$. Its functional form is
\beql{sat.scale}
\Q_s^2= A^{1/3} x_0^\lambda\, e^{\lambda y}\,s^{\lambda/2}\,\mathrm{GeV}^2\,,
\eeq
where $s$ is the square of the center-of-mass energy and $y$ is rapidity with respect to the central rapidity. The anomalous dimension is 
\beql{anom.dim}
\gamma = \gamma_s+\frac{1}{c\, \lambda \,(\ln\surd s+y)}\ln \left( \frac{2}{r\Q_s}\right)\,.
\eeq
The gluon dipole scattering amplitude can be calculated using \eq{Ng-N}.
Parameters $\gamma_s=0.628$ and $c=9.9$ follow from the BFKL dynamics \cite{Iancu:2002tr}, while $\mathcal{N}_0=0.7$ and $\lambda=0.28$ are fitted to the DIS data. Constants $a$ and $b$ are uniquely fixed from by the requirement of continuity of the amplitude and its first derivative.

%%%%%%%%%%%%%%%%%%%%%%%%%%%%%%%%%%%%%


\begin{thebibliography}{80}

%----  CGC  ---------

%\cite{Gribov:1984tu}
\bibitem{Gribov:1984tu} 
  L.~V.~Gribov, E.~M.~Levin and M.~G.~Ryskin,
  %``Semihard Processes in QCD,''
  Phys.\ Rept.\  {\bf 100}, 1 (1983).
  %%CITATION = PRPLC,100,1;%%
  
\bibitem{LR}
E.M.  Levin and M.G Ryskin, Nucl. Phys. {\bf B304}, 805 (1988);
Sov. J.  Nucl. Phys. {\bf 45}, 150 (1987); {\bf 41}, 300 (1985).

%\cite{Mueller:1985wy}
\bibitem{Mueller:1985wy} 
  A.~H.~Mueller and J.~-w.~Qiu,
  %``Gluon Recombination and Shadowing at Small Values of x,''
  Nucl.\ Phys.\ B {\bf 268}, 427 (1986).
  %%CITATION = NUPHA,B268,427;%%
  
  %\cite{Mueller:1993rr}
\bibitem{Mueller:1993rr} 
  A.~H.~Mueller,
  %``Soft gluons in the infinite momentum wave function and the BFKL pomeron,''
  Nucl.\ Phys.\ B {\bf 415}, 373 (1994).
  %%CITATION = NUPHA,B415,373;%%
  
  %\cite{Mueller:1994jq}
\bibitem{Mueller:1994jq} 
  A.~H.~Mueller and B.~Patel,
  %``Single and double BFKL pomeron exchange and a dipole picture of high-energy hard processes,''
  Nucl.\ Phys.\ B {\bf 425}, 471 (1994)
  [hep-ph/9403256].
  %%CITATION = HEP-PH/9403256;%%
  
  %\cite{Mueller:1994gb}
\bibitem{Mueller:1994gb} 
  A.~H.~Mueller,
  %``Unitarity and the BFKL pomeron,''
  Nucl.\ Phys.\ B {\bf 437}, 107 (1995)
  [hep-ph/9408245].
  %%CITATION = HEP-PH/9408245;%%
  
  
  %\cite{McLerran:1993ka}
\bibitem{McLerran:1993ka}
  L.~D.~McLerran and R.~Venugopalan,
  %``Gluon distribution functions for very large nuclei at small transverse
  %momentum,''
  Phys.\ Rev.\  D {\bf 49}, 3352 (1994)
  [arXiv:hep-ph/9311205].
  %%CITATION = PHRVA,D49,3352;%%
  
%\cite{McLerran:1993ni}
\bibitem{McLerran:1993ni}
  L.~D.~McLerran and R.~Venugopalan,
  %``Computing quark and gluon distribution functions for very large nuclei,''
  Phys.\ Rev.\  D {\bf 49}, 2233 (1994)
  [arXiv:hep-ph/9309289].
  %%CITATION = PHRVA,D49,2233;%%

%\cite{McLerran:1994vd}
\bibitem{McLerran:1994vd}
  L.~D.~McLerran and R.~Venugopalan,
  %``Green's functions in the color field of a large nucleus,''
  Phys.\ Rev.\  D {\bf 50} (1994) 2225
  [arXiv:hep-ph/9402335].
  %%CITATION = PHRVA,D50,2225;%%
  
  %\cite{Kovchegov:1996ty}
\bibitem{Kovchegov:1996ty}
  Y.~V.~Kovchegov,
  %``Non-Abelian Weizsaecker-Williams field and a two-dimensional effective
  %color charge density for a very large nucleus,''
  Phys.\ Rev.\  D {\bf 54}, 5463 (1996)
  [arXiv:hep-ph/9605446].
  %%CITATION = PHRVA,D54,5463;%%
  
  %\cite{Kovchegov:1997pc}
\bibitem{Kovchegov:1997pc}
  Y.~V.~Kovchegov,
  %``Quantum structure of the non-Abelian Weizsaecker-Williams field for a  very
  %large nucleus,''
  Phys.\ Rev.\  D {\bf 55}, 5445 (1997)
  [arXiv:hep-ph/9701229].
  %%CITATION = PHRVA,D55,5445;%%
  
  %\cite{JalilianMarian:1996xn}
\bibitem{JalilianMarian:1996xn}
  J.~Jalilian-Marian, A.~Kovner, L.~D.~McLerran and H.~Weigert,
  %``The intrinsic glue distribution at very small x,''
  Phys.\ Rev.\  D {\bf 55}, 5414 (1997)
  [arXiv:hep-ph/9606337].
  %%CITATION = PHRVA,D55,5414;%%
  
  %\cite{JalilianMarian:1997jx}
\bibitem{JalilianMarian:1997jx}
  J.~Jalilian-Marian, A.~Kovner, A.~Leonidov and H.~Weigert,
  %``The BFKL equation from the Wilson renormalization group,''
  Nucl.\ Phys.\  B {\bf 504} (1997) 415
  [arXiv:hep-ph/9701284].
  %%CITATION = NUPHA,B504,415;%%
  
  %\cite{JalilianMarian:1997gr}
\bibitem{JalilianMarian:1997gr}
  J.~Jalilian-Marian, A.~Kovner, A.~Leonidov and H.~Weigert,
  %``The Wilson renormalization group for low x physics: Towards the high
  %density regime,''
  Phys.\ Rev.\  D {\bf 59}, 014014 (1999)
  [arXiv:hep-ph/9706377].
  %%CITATION = PHRVA,D59,014014;%%
  
  %\cite{JalilianMarian:1997dw}
\bibitem{JalilianMarian:1997dw}
  J.~Jalilian-Marian, A.~Kovner and H.~Weigert,
  %``The Wilson renormalization group for low x physics: Gluon evolution at
  %finite parton density,''
  Phys.\ Rev.\  D {\bf 59}, 014015 (1999)
  [arXiv:hep-ph/9709432].
  %%CITATION = PHRVA,D59,014015;%%
  
  %\cite{JalilianMarian:1998cb}
\bibitem{JalilianMarian:1998cb}
  J.~Jalilian-Marian, A.~Kovner, A.~Leonidov and H.~Weigert,
  %``Unitarization of gluon distribution in the doubly logarithmic regime at
  %high density,''
  Phys.\ Rev.\  D {\bf 59}, 034007 (1999)
  [Erratum-ibid.\  D {\bf 59}, 099903 (1999)]
  [arXiv:hep-ph/9807462].
  %%CITATION = PHRVA,D59,034007;%%
  
  %\cite{Kovner:2000pt}
\bibitem{Kovner:2000pt}
  A.~Kovner, J.~G.~Milhano and H.~Weigert,
  %``Relating different approaches to nonlinear QCD evolution at finite  gluon
  %density,''
  Phys.\ Rev.\  D {\bf 62}, 114005 (2000)
  [arXiv:hep-ph/0004014].
  %%CITATION = PHRVA,D62,114005;%%
  
  %\cite{Weigert:2000gi}
\bibitem{Weigert:2000gi}
  H.~Weigert,
  %``Unitarity at small Bjorken x,''
  Nucl.\ Phys.\  A {\bf 703}, 823 (2002)
  [arXiv:hep-ph/0004044].
  %%CITATION = NUPHA,A703,823;%%
  
  %\cite{Iancu:2000hn}
\bibitem{Iancu:2000hn}
  E.~Iancu, A.~Leonidov and L.~D.~McLerran,
  %``Nonlinear gluon evolution in the color glass condensate. I,''
  Nucl.\ Phys.\  A {\bf 692}, 583 (2001)
  [arXiv:hep-ph/0011241].
  %%CITATION = NUPHA,A692,583;%%
  
  
  %\cite{Ferreiro:2001qy}
\bibitem{Ferreiro:2001qy}
  E.~Ferreiro, E.~Iancu, A.~Leonidov and L.~McLerran,
  %``Nonlinear gluon evolution in the color glass condensate. II,''
  Nucl.\ Phys.\  A {\bf 703}, 489 (2002)
  [arXiv:hep-ph/0109115].
  %%CITATION = NUPHA,A703,489;%%
  
  %\cite{Kovchegov:1999yj}
\bibitem{Kovchegov:1999yj}
  Y.~V.~Kovchegov,
  %``Small-x F2 structure function of a nucleus including multiple pomeron
  %exchanges,''
  Phys.\ Rev.\  D {\bf 60}, 034008 (1999)
  [arXiv:hep-ph/9901281].
  %%CITATION = PHRVA,D60,034008;%%
  
  %\cite{Kovchegov:1999ua}
\bibitem{Kovchegov:1999ua}
  Y.~V.~Kovchegov,
  %``Unitarization of the BFKL pomeron on a nucleus,''
  Phys.\ Rev.\  D {\bf 61}, 074018 (2000)
  [arXiv:hep-ph/9905214].
  %%CITATION = PHRVA,D61,074018;%%
  
  %\cite{Balitsky:1995ub}
\bibitem{Balitsky:1995ub}
  I.~Balitsky,
  %``Operator expansion for high-energy scattering,''
  Nucl.\ Phys.\  B {\bf 463}, 99 (1996)
  [arXiv:hep-ph/9509348].
  %%CITATION = NUPHA,B463,99;%%
  
  %\cite{Balitsky:1997mk}
\bibitem{Balitsky:1997mk}
  I.~Balitsky,
  %``Operator expansion for diffractive high-energy scattering,''
  arXiv:hep-ph/9706411.
  %%CITATION = HEP-PH/9706411;%%
  
  %\cite{Balitsky:1998ya}
\bibitem{Balitsky:1998ya}
  I.~Balitsky,
  %``Factorization and high-energy effective action,''
  Phys.\ Rev.\  D {\bf 60}, 014020 (1999)
  [arXiv:hep-ph/9812311].
  %%CITATION = PHRVA,D60,014020;%%
  
  %\cite{Iancu:2003xm}
\bibitem{Iancu:2003xm}
  E.~Iancu and R.~Venugopalan,
  %``The color glass condensate and high energy scattering in QCD,''
  arXiv:hep-ph/0303204.
  %%CITATION = HEP-PH/0303204;%%
  
  %\cite{JalilianMarian:2005jf}
\bibitem{JalilianMarian:2005jf}
  J.~Jalilian-Marian and Y.~V.~Kovchegov,
  %``Saturation physics and deuteron gold collisions at RHIC,''
  Prog.\ Part.\ Nucl.\ Phys.\  {\bf 56}, 104 (2006)
  [arXiv:hep-ph/0505052].
  %%CITATION = PPNPD,56,104;%%
  
  %-------------------------------------------
  
  %-------------- Inclusive gluons -------------

%\cite{Kovchegov:1998bi}
\bibitem{Kovchegov:1998bi}
  Y.~V.~Kovchegov and A.~H.~Mueller,
%   ``Gluon production in current nucleus and nucleon nucleus collisions in  a
  %quasi-classical approximation,''
  Nucl.\ Phys.\ B {\bf 529}, 451 (1998).
  %%CITATION = HEP-PH 9802440;%%

%\cite{Kovchegov:2001sc}
\bibitem{Kovchegov:2001sc}
  Y.~V.~Kovchegov and K.~Tuchin,
  %``Inclusive gluon production in DIS at high parton density,''
  Phys.\ Rev.\ D {\bf 65}, 074026 (2002).
  %%CITATION = HEP-PH 0111362;%%


%\cite{Braun:2000bh}
\bibitem{Braun:2000bh}
  M.~A.~Braun,
  % ``Inclusive jet production on the nucleus in the perturbative QCD with  N(c)
  %--> infinity,''
  Phys.\ Lett.\ B {\bf 483}, 105 (2000)
  [arXiv:hep-ph/0003003].
  %%CITATION = HEP-PH 0003003;%%

%\cite{Dumitru:2001ux}
\bibitem{Dumitru:2001ux}
  A.~Dumitru and L.~D.~McLerran,
  %``How protons shatter colored glass,''
  Nucl.\ Phys.\ A {\bf 700}, 492 (2002)
  [arXiv:hep-ph/0105268].
  %%CITATION = HEP-PH 0105268;%%

%\cite{Blaizot:2004wu}
\bibitem{Blaizot:2004wu}
  J.~P.~Blaizot, F.~Gelis, and R.~Venugopalan,
 %  ``High energy p A collisions in the color glass condensate approach. I:
  %Gluon production and the Cronin effect,''
  Nucl.\ Phys.\ A {\bf 743}, 13 (2004)
  [arXiv:hep-ph/0402256].
  %%CITATION = HEP-PH 0402256;%%

%\cite{Kharzeev:2002pc}
\bibitem{Kharzeev:2002pc}
  D.~Kharzeev, E.~Levin, and L.~McLerran,
  %``Parton saturation and N(part) scaling of semi-hard processes in QCD,''
  Phys.\ Lett.\ B {\bf 561}, 93 (2003)
  [arXiv:hep-ph/0210332].
  %%CITATION = HEP-PH 0210332;%%

%\cite{Kharzeev:2003wz}
\bibitem{Kharzeev:2003wz}
  D.~Kharzeev, Y.~V.~Kovchegov, and K.~Tuchin,
  %``Cronin effect and high-p(T) suppression in p A collisions,''
  Phys.\ Rev.\ D {\bf 68}, 094013 (2003)
  [arXiv:hep-ph/0307037].
  %%CITATION = HEP-PH 0307037;%%

%\cite{Kharzeev:2004yx}
\bibitem{Kharzeev:2004yx}
  D.~Kharzeev, Y.~V.~Kovchegov, and K.~Tuchin,
  %``Nuclear modification factor in d + Au collisions: Onset of suppression  in
  %the color glass condensate,''
  Phys.\ Lett.\  B {\bf 599}, 23 (2004)
  [arXiv:hep-ph/0405045].
  %%CITATION = PHLTA,B599,23;%%

%\cite{Baier:2003hr}
\bibitem{Baier:2003hr}
  R.~Baier, A.~Kovner, and U.~A.~Wiedemann,
 %  ``Saturation and parton level Cronin effect: Enhancement vs suppression  of
  %gluon production in p A and A A collisions,''
  Phys.\ Rev.\ D {\bf 68}, 054009 (2003)
  [arXiv:hep-ph/0305265].
  %%CITATION = HEP-PH 0305265;%%



%\cite{Iancu:2004bx}
\bibitem{Iancu:2004bx}
  E.~Iancu, K.~Itakura, and D.~N.~Triantafyllopoulos,
 %  ``Cronin effect and high-p(T) suppression in the nuclear gluon  distribution
  %at small x,''
  Nucl.\ Phys.\ A {\bf 742}, 182 (2004)
  [arXiv:hep-ph/0403103].
  %%CITATION = HEP-PH 0403103;%%
%------------------------------------------

%\cite{Mueller:1989st}
\bibitem{Mueller:1989st}
  A.~H.~Mueller,
  %``Small x Behavior and Parton Saturation: A QCD Model,''
  Nucl.\ Phys.\  B {\bf 335}, 115 (1990).
  %%CITATION = NUPHA,B335,115;%%
  
  %--------   BFKL ----------
  
%\cite{Kuraev:1977fs}
\bibitem{Kuraev:1977fs}
  E.~A.~Kuraev, L.~N.~Lipatov, and V.~S.~Fadin,
  %``The Pomeranchuk Singularity In Nonabelian Gauge Theories,''
  Sov.\ Phys.\ JETP {\bf 45}, 199 (1977)
  [Zh.\ Eksp.\ Teor.\ Fiz.\  {\bf 72}, 377 (1977)].
  %%CITATION = ZETFA,72,377;%%

%\cite{Balitsky:1978ic}
\bibitem{Balitsky:1978ic}
  I.~I.~Balitsky and L.~N.~Lipatov,
  %``The Pomeranchuk Singularity In Quantum Chromodynamics,''
  Sov.\ J.\ Nucl.\ Phys.\  {\bf 28} (1978) 822
  [Yad.\ Fiz.\  {\bf 28} (1978) 1597].
  %%CITATION = YAFIA,28,1597;%%
  
 %-=------------------------------
 
 %--------   NLO BFKL ----------
  
  %\cite{Fadin:1998py}
\bibitem{Fadin:1998py} 
  V.~S.~Fadin and L.~N.~Lipatov,
  %``BFKL pomeron in the next-to-leading approximation,''
  Phys.\ Lett.\ B {\bf 429}, 127 (1998)
  [hep-ph/9802290].
  %%CITATION = HEP-PH/9802290;%%
  
  %\cite{Ciafaloni:1998gs}
\bibitem{Ciafaloni:1998gs} 
  M.~Ciafaloni and G.~Camici,
  %``Energy scale(s) and next-to-leading BFKL equation,''
  Phys.\ Lett.\ B {\bf 430}, 349 (1998)
  [hep-ph/9803389].
  %%CITATION = HEP-PH/9803389;%%
  
  %\cite{Ciafaloni:2003rd}
\bibitem{Ciafaloni:2003rd} 
  M.~Ciafaloni, D.~Colferai, G.~P.~Salam and A.~M.~Stasto,
  %``Renormalization group improved small x Green's function,''
  Phys.\ Rev.\ D {\bf 68}, 114003 (2003)
  [hep-ph/0307188].
  %%CITATION = HEP-PH/0307188;%%
  
%\cite{Ciafaloni:2002xf}
\bibitem{Ciafaloni:2002xf} 
  M.~Ciafaloni, D.~Colferai, G.~P.~Salam and A.~M.~Stasto,
  %``Expanding running coupling effects in the hard pomeron,''
  Phys.\ Rev.\ D {\bf 66}, 054014 (2002)
  [hep-ph/0204282].
  %%CITATION = HEP-PH/0204282;%%
  
  %\cite{Ciafaloni:2002xk}
\bibitem{Ciafaloni:2002xk} 
  M.~Ciafaloni, D.~Colferai, G.~P.~Salam and A.~M.~Stasto,
  %``Tunneling transition to the pomeron regime,''
  Phys.\ Lett.\ B {\bf 541}, 314 (2002)
  [hep-ph/0204287].
  %%CITATION = HEP-PH/0204287;%%
  
  %\cite{Forshaw:2000hv}
\bibitem{Forshaw:2000hv} 
  J.~R.~Forshaw, D.~A.~Ross and A.~Sabio Vera,
  %``Solving the BFKL equation with running coupling,''
  Phys.\ Lett.\ B {\bf 498}, 149 (2001)
  [hep-ph/0011047].
  %%CITATION = HEP-PH/0011047;%%
  
  %\cite{Ciafaloni:2001db}
\bibitem{Ciafaloni:2001db} 
  M.~Ciafaloni, M.~Taiuti and A.~H.~Mueller,
  %``Diffusion corrections to the hard pomeron,''
  Nucl.\ Phys.\ B {\bf 616}, 349 (2001)
  [hep-ph/0107009].
  %%CITATION = HEP-PH/0107009;%%
  
  %\cite{Brodsky:1998kn}
\bibitem{Brodsky:1998kn} 
  S.~J.~Brodsky, V.~S.~Fadin, V.~T.~Kim, L.~N.~Lipatov and G.~B.~Pivovarov,
  %``The QCD pomeron with optimal renormalization,''
  JETP Lett.\  {\bf 70}, 155 (1999)
  [hep-ph/9901229].
  %%CITATION = HEP-PH/9901229;%%
  

%\cite{Ross:1998xw}
\bibitem{Ross:1998xw} 
  D.~A.~Ross,
  %``The Effect of higher order corrections to the BFKL equation on the perturbative pomeron,''
  Phys.\ Lett.\ B {\bf 431}, 161 (1998)
  [hep-ph/9804332].
  %%CITATION = HEP-PH/9804332;%%
  
  %\cite{Levin:1998pka}
\bibitem{Levin:1998pka} 
  E.~Levin,
  %``The BFKL high-energy asymptotics in the next-to-leading approximation,''
  hep-ph/9806228.
  %%CITATION = HEP-PH/9806228;%%
  
  %\cite{Armesto:1998gt}
\bibitem{Armesto:1998gt} 
  N.~Armesto, J.~Bartels and M.~A.~Braun,
  %``On the second order corrections to the hard pomeron and the running coupling,''
  Phys.\ Lett.\ B {\bf 442}, 459 (1998)
  [hep-ph/9808340].
  %%CITATION = HEP-PH/9808340;%%
  
  %\cite{Kovchegov:1998ae}
\bibitem{Kovchegov:1998ae} 
  Y.~V.~Kovchegov and A.~H.~Mueller,
  %``Running coupling effects in BFKL evolution,''
  Phys.\ Lett.\ B {\bf 439}, 428 (1998)
  [hep-ph/9805208].
  %%CITATION = HEP-PH/9805208;%%
  
 %--------------------------------
 
 
  
 %------  Running coupling corrections to BK -------- 
  
  %\cite{Levin:1994di}
\bibitem{Levin:1994di} 
  E.~Levin,
  %``Renormalons at low x,''
  Nucl.\ Phys.\ B {\bf 453}, 303 (1995)
  [hep-ph/9412345].
  %%CITATION = HEP-PH/9412345;%%
  
  %\cite{Braun:1994mw}
\bibitem{Braun:1994mw} 
  M.~A.~Braun,
  %``Reggeized gluons with a running coupling constant,''
  Phys.\ Lett.\ B {\bf 348}, 190 (1995)
  [hep-ph/9408261].
  %%CITATION = HEP-PH/9408261;%%
  
  %\cite{Kovchegov:2006vj}
\bibitem{Kovchegov:2006vj} 
  Y.~V.~Kovchegov and H.~Weigert,
  %``Triumvirate of Running Couplings in Small-x Evolution,''
  Nucl.\ Phys.\ A {\bf 784}, 188 (2007)
  [hep-ph/0609090].
  %%CITATION = HEP-PH/0609090;%%


%\cite{Kovchegov:2006wf}
\bibitem{Kovchegov:2006wf} 
  Y.~V.~Kovchegov and H.~Weigert,
  %``Quark loop contribution to BFKL evolution: Running coupling and leading-N(f) NLO intercept,''
  Nucl.\ Phys.\ A {\bf 789}, 260 (2007)
  [hep-ph/0612071].
  %%CITATION = HEP-PH/0612071;%%

%\cite{Kovchegov:2007vf}
\bibitem{Kovchegov:2007vf} 
  Y.~V.~Kovchegov and H.~Weigert,
  %``Collinear Singularities and Running Coupling Corrections to Gluon Production in CGC,''
  Nucl.\ Phys.\ A {\bf 807}, 158 (2008)
  [arXiv:0712.3732 [hep-ph]].
  %%CITATION = ARXIV:0712.3732;%%
  
  
  %\cite{Balitsky:2006wa}
\bibitem{Balitsky:2006wa} 
  I.~Balitsky,
  %``Quark contribution to the small-x evolution of color dipole,''
  Phys.\ Rev.\ D {\bf 75}, 014001 (2007)
  [hep-ph/0609105].
  %%CITATION = HEP-PH/0609105;%%

%------------------------------------------------

  %\cite{Mueller:2002zm}
\bibitem{Mueller:2002zm}
  A.~H.~Mueller, D.~N.~Triantafyllopoulos,
  %``The Energy dependence of the saturation momentum,''
  Nucl.\ Phys.\  {\bf B640}, 331-350 (2002).
  [hep-ph/0205167].
  
  %\cite{Triantafyllopoulos:2002nz}
\bibitem{Triantafyllopoulos:2002nz} 
  D.~N.~Triantafyllopoulos,
  %``The Energy dependence of the saturation momentum from RG improved BFKL evolution,''
  Nucl.\ Phys.\ B {\bf 648}, 293 (2003)
  [hep-ph/0209121].
  %%CITATION = HEP-PH/0209121;%%


%\cite{Kuokkanen:2011}
\bibitem{Kuokkanen:2011}
  J.~Kuokkanen, K.~Rummukainen, H.~Weigert,
  %``HERA-data in the light of small x evolution with state of the art NLO input,''
  arXiv:1108.1867 [hep-ph].
  %%CITATION = ARXIV:1108.1867;%%
  
  %\cite{Weigert:2007hk}
\bibitem{Weigert:2007hk} 
  H.~Weigert,
  %``A compact introduction to evolution at small x and the color glass condensate,''
  Nucl.\ Phys.\ A {\bf 783}, 165 (2007).
  %%CITATION = NUPHA,A783,165;%%
  
  %\cite{Chachamis:2004ab}
\bibitem{Chachamis:2004ab} 
  G.~Chachamis, M.~Lublinsky and A.~Sabio Vera,
  %``Higher order effects in non linear evolution from a veto in rapidities,''
  Nucl.\ Phys.\ A {\bf 748}, 649 (2005)
  [hep-ph/0408333].
  %%CITATION = HEP-PH/0408333;%%

%\cite{Kormilitzin:2010at}
\bibitem{Kormilitzin:2010at}
  A.~Kormilitzin and E.~Levin,
  %``Non-linear equation: energy conservation and impact parameter dependence,''
  arXiv:1009.1468 [hep-ph].
  %%CITATION = ARXIV:1009.1468;%%
  
  %\cite{Gotsman:2004xb}
\bibitem{Gotsman:2004xb}
  E.~Gotsman, E.~Levin, U.~Maor and E.~Naftali,
  %``A Modified Balitsky-Kovchegov equation,''
  Nucl.\ Phys.\  A {\bf 750}, 391 (2005)
  [arXiv:hep-ph/0411242].
  %%CITATION = NUPHA,A750,391;%%

%\cite{Gotsman:2002yy}
\bibitem{Gotsman:2002yy} 
  E.~Gotsman, E.~Levin, M.~Lublinsky and U.~Maor,
  %``Towards a new global QCD analysis: Low x DIS data from nonlinear evolution,''
  Eur.\ Phys.\ J.\ C {\bf 27}, 411 (2003)
  [hep-ph/0209074].
  %%CITATION = HEP-PH/0209074;%%

%\cite{Kowalski:2006hc}
\bibitem{Kowalski:2006hc}
  H.~Kowalski, L.~Motyka and G.~Watt,
  %``Exclusive diffractive processes at HERA within the dipole picture,''
  Phys.\ Rev.\  D {\bf 74}, 074016 (2006)
  [arXiv:hep-ph/0606272].
  %%CITATION = PHRVA,D74,074016;%%
  
  
  %\cite{Nikolaev:1990ja}
\bibitem{Nikolaev:1990ja} 
  N.~N.~Nikolaev and B.~G.~Zakharov,
  %``Color transparency and scaling properties of nuclear shadowing in deep inelastic scattering,''
  Z.\ Phys.\ C {\bf 49}, 607 (1991).
  %%CITATION = ZEPYA,C49,607;%%
  
  %\cite{Li:2008bm}
\bibitem{Li:2008bm} 
  Y.~Li and K.~Tuchin,
  %``Gluon multiplicity in coherent diffraction of dipole on a heavy nucleus,''
  Phys.\ Rev.\ D {\bf 77}, 114012 (2008)
  [arXiv:0802.2954 [hep-ph]].
  %%CITATION = ARXIV:0802.2954;%%

%\cite{Li:2008jz}
\bibitem{Li:2008jz}
  Y.~Li and K.~Tuchin,
  %``Spectrum of diffractively produced gluons in dipole-nucleus collisions,''
  arXiv:0803.1608 [hep-ph].
  %%CITATION = ARXIV:0803.1608;%%
  
  %\cite{Li:2008se}
\bibitem{Li:2008se}
  Y.~Li and K.~Tuchin,
  %``Probing the low-x structure of nuclear matter with diffractive hadron
  %production in pA collisions,''
  Phys.\ Rev.\  C {\bf 78}, 024905 (2008)
  [arXiv:0806.2087 [hep-ph]].
  %%CITATION = PHRVA,C78,024905;%%
  
%\cite{Levin:1999mw}
\bibitem{Levin:1999mw} 
  E.~Levin and K.~Tuchin,
  %``Solution to the evolution equation for high parton density QCD,''
  Nucl.\ Phys.\ B {\bf 573}, 833 (2000)
  [hep-ph/9908317].
  %%CITATION = HEP-PH/9908317;%%
  
%\cite{Levin:2000mv}
\bibitem{Levin:2000mv}
  E.~Levin and K.~Tuchin,
  %``New Scaling at High Energy DIS,''
  Nucl.\ Phys.\  A {\bf 691}, 779 (2001)
  [arXiv:hep-ph/0012167].
  %%CITATION = NUPHA,A691,779;%%

%\cite{Levin:2001cv}
\bibitem{Levin:2001cv} 
  E.~Levin and K.~Tuchin,
  %``Nonlinear evolution and saturation for heavy nuclei in DIS,''
  Nucl.\ Phys.\ A {\bf 693}, 787 (2001)
  [hep-ph/0101275].
  %%CITATION = HEP-PH/0101275;%%


%--------------------------------------------------------------

  %\cite{Ellis:1993rb}
\bibitem{Ellis:1993rb}
  R.~K.~Ellis, Z.~Kunszt and E.~M.~Levin,
  %``The Evolution Of Parton Distributions At Small X,''
  Nucl.\ Phys.\  B {\bf 420}, 517 (1994)
  [Erratum-ibid.\  B {\bf 433}, 498 (1995)].
  %%CITATION = NUPHA,B420,517;%%

%\cite{AyalaFilho:1997du}
\bibitem{AyalaFilho:1997du}
  A.~L.~Ayala Filho, M.~B.~Gay Ducati and E.~M.~Levin,
  %``Parton densities in a nucleon,''
  Nucl.\ Phys.\  B {\bf 511}, 355 (1998)
  [arXiv:hep-ph/9706347].
  %%CITATION = NUPHA,B511,355;%%


%\cite{Marquet:2007xx}
\bibitem{Marquet:2007xx}
  C.~Marquet and C.~Royon,
  %``Azimuthal decorrelation of Mueller-Navelet jets at the Tevatron and the
  %LHC,''
  Phys.\ Rev.\  D {\bf 79}, 034028 (2009)
  [arXiv:0704.3409 [hep-ph]].
  %%CITATION = PHRVA,D79,034028;%%


  
    %\cite{Tuchin:2007pf}
\bibitem{Tuchin:2007pf}
  K.~Tuchin,
  %``Forward hadron production in high energy p A collisions: From RHIC to
  %LHC,''
  Nucl.\ Phys.\  A {\bf 798}, 61 (2008)
  [arXiv:0705.2193 [hep-ph]].
  %%CITATION = NUPHA,A798,61;%%
  
  %\cite{Iancu:2002tr}
\bibitem{Iancu:2002tr}
  E.~Iancu, K.~Itakura and L.~McLerran,
  %``Geometric scaling above the saturation scale,''
  Nucl.\ Phys.\  A {\bf 708}, 327 (2002)
  [arXiv:hep-ph/0203137].
  %%CITATION = NUPHA,A708,327;%%
  
%\cite{Kniehl:2000hk}
\bibitem{Kniehl:2000hk} 
  B.~A.~Kniehl, G.~Kramer and B.~Potter,
  %``Testing the universality of fragmentation functions,''
  Nucl.\ Phys.\ B {\bf 597}, 337 (2001)
  [hep-ph/0011155].
  %%CITATION = HEP-PH/0011155;%%
  
  %\cite{Boer:2011fh}
\bibitem{Boer:2011fh} 
  D.~Boer, M.~Diehl, R.~Milner, R.~Venugopalan, W.~Vogelsang, D.~Kaplan, H.~Montgomery and S.~Vigdor {\it et al.},
  %``Gluons and the quark sea at high energies: Distributions, polarization, tomography,''
  arXiv:1108.1713 [nucl-th].
  %%CITATION = ARXIV:1108.1713;%%
  
\end{thebibliography}
\end{document}